\documentclass[12]{article}
\usepackage{graphics,graphicx,amsmath,amssymb}
\usepackage{color,xcolor,comment}

\newcommand{\II}{{\boldmath \mbox{$I$}}}

\newcommand{\JJ}{{\boldmath \mbox{$J$}}}

\newcommand{\xx}{{\boldmath \mbox{$x$}}}

\newcommand{\rr}{{\boldmath \mbox{$r$}}}
\newcommand{\mm}{{\boldmath \mbox{$m$}}}

\newcommand{\pp}{{\boldmath \mbox{$p$}}}

\newcommand{\nn}{{\boldmath \mbox{$n$}}}

\newcommand{\zzeta}{{\boldmath\mbox{$\zeta$}}}

\newcommand{\nnu}{{\boldmath\mbox{$\nu$}}}

\newlength{\defbaselineskip}
\newcommand{\setlinespacing}[1]%
           {\setlength{\baselineskip}{#1 \defbaselineskip}}

\title{\textbf{Fluctuating Multiscale Mass  Action Law}}
\author{
Abdellah Ajji $^{1}$,  Jamal Chaouki $^{1}$,  Miroslav Grmela $^{1}$
\footnote{corresponding author: e-mail:
miroslav.grmela@polymtl.ca},  \\  V\'{a}clav Klika $^{2}$,  Michal Pavelka$^{3},$
\vspace {0.5cm}\\
$^1$ G\'{e}nie chimique, \'{E}cole Polytechnique de Montr\'{e}al,\\
  C.P.6079 suc. Centre-ville,
 Montr\'{e}al, H3C 3A7,  Qu\'{e}bec, Canada \\
  $^2$  Czech Technical University in Prague, Department of Mathematics -- FNSPE,\\ Trojanova 13, 120 00 Prague, Czech Republic \\
 $^3$ Mathematical Institute, Faculty of Mathematics, Charles University,\\ Prague,
 Sokolovsk\'{a} 83, 18675 Prague, Czech Republic\\
Czech Republic}

 \date{}

\begin{document}

\maketitle

\begin{abstract}

The classical mass action law in chemical kinetics is put into the context of multiscale thermodynamics.
Despite the purely dissipative character of the classical mass action law, inertial effects also play a role in chemical kinetics. Therefore, the kinetics is extended to an enlarged state space with reaction rates as new state variables and bringing inertial effects. 
The dynamics is then lifted to the Liouville description within kinetic theory on the enlarged state space in order to include fluctuations. 
Subsequently, the kinetic theory is reduced to its first and second moments, leading to a new Grad-like hierarchy in chemical kinetics, expressing the multiscale nature of the chemical kinetic with inertia.
Dissipation within the extended state space is proposed and it is shown to lead to the classical mass action law when the moments relax to their respective quasi-equilibria. In particular, we demonstrate the possibility of oscillating homogeneous chemical reactions and how the correlations of fluctuations correspond with the chemical kinetics.
\end{abstract}


\section{Introduction}\label{Int}

The classical mass action law (classical MAL) \cite{guldberg1879} provides a framework for investigating the time evolution of chemically  reacting systems.  The extended mass action law (extended MAL) introduced in this paper  allows  to include the  influence of fast reactions involving intermediate components, the  influence of external driving forces, and the  influence of fluctuations. The compatibility of the extended and classical theories is proven by comparing solutions to their governing equations. A theory  involving more details (called hereafter upper level theory) is compatible with a theory involving less detail (called hereafter lower level theory) if the lower phase portrait emerges as a pattern in the upper phase portrait. An upper (resp. lower) phase portrait is a collection of solutions passing through all points in the upper (resp. lower) state space for an ensemble of macroscopic systems.
 In particular, the time evolution is compatible with the equilibrium thermodynamics if it describes faithfully the experimentally observed approach to equilibrium states at which the observed behavior is well described by the classical equilibrium thermodynamics.

If external forces  prevent the approach to equilibrium, the equilibrium thermodynamics does not apply. In such a situation, the approach from an upper to the lower level on which the experimentally observed behavior of the externally driven system is found to be well described, provides a basis for another thermodynamics, called rate thermodynamics \cite{MG-CR}, \cite{miroslav-MT} in which  the state space is the space of  forces and fluxes, see e.g. \cite{renger-fluxes} for a connection with the principle of large deviations. The extended mass-action-law theories  introduced in this paper make possible to formulate such a theory for chemically reacting systems.

Beside the need to enlarge the domain of applicability of the classical MAL, our principal motivation is to illustrate the viewpoint and the methods of multiscale thermodynamics \cite{adv}, \cite{miroslav-guide}, \cite{pkg}, \cite{miroslav-ptrsa}. Specific illustrations of the earlier versions of the extended mass action law \cite{miroslav-driven},\cite{grchemkin}  have been worked out in \cite{klikajpcb2009coupling,klika2013-pre} (coupling of chemical reactions with mechanics occurring for instance in bone  growth \cite{vasek-remodelling,klikajpcb2009coupling})  and in \cite{vk2014-bz}  (using chemical reactions to produce  mechanical work).

In Section \ref{MAL} we formulate an extended MAL and reduce it (by investigating its solutions in the following section) to  the classical Guldberg-Waage mass action law. In Section \ref{KMAL} we lift extended MAL to kinetic theory, similarly as \cite{kramers}, allowing to include fluctuations into the analysis, but including also the momentum-like dimensions, as e.g. in \cite{Cafel}, to describe electron transfer with inertial effects. In Sec. \ref{2nd} we introduce dissipation into the kinetic theory (forming a Fokker-Planck equation) and, subsequently, reduce the description to moments of the distribution function. In other words, we lift the Guldberg-Waage dynamics to a higher level of description, introducing the inertia and dynamics of distribution functions. Subsequently, we reduce the description to a lower level keeping the reminiscence of the higher level, which leads to a novel Grad-like hierarchy of chemical kinetics with inertia. The novelty of this paper lies in the geometric formulation of chemical reactions with inertia, in the subsequent multiscale analysis, and finally in the construction of the Grad-like hierarchy of moments for chemical kinetics.

\section{Mass Action Law within multiscale thermodynamics} \label{clasMAL}

Let $p$ components $\mathbb{A}_1,...,\mathbb{A}_p$ undergo $q$ chemical reactions
\begin{equation}\label{reac}
\mu_{11}\mathbb{A}_1+...+\mu_{p1}\mathbb{A}_p\leftrightarrows \nu_{11}\mathbb{A}_1+...+\nu_{p1}\mathbb{A}_p,
\end{equation}
where
\begin{equation}\label{gamma}
\gamma_{\alpha j}=\nu_{\alpha j}-\mu_{\alpha j};\,\,\,\alpha=1,...,p;\,\,\,j=1,...,q,
\end{equation}
are the stoichiometric coefficients and
\begin{equation}\label{Gamma}
\gamma=\left(\begin{array}{ccc}\gamma_{11}&\cdots&\gamma_{p1}\\ \vdots&\vdots&\vdots\\ \gamma_{1q}&\cdots&\gamma_{pq}\end{array}\right),
\end{equation}
the stoichiometric matrix. Hereafter, we use the lowercase Greek  letters ($\alpha, \beta, \epsilon= 1,...,p$) to label the components and the lowercase Roman letters ($i,j,k,l=1,...,q$) to label the  reactions. We shall also use the summation convention over  repeated indices.

The time evolution of the number of moles 
$\nn=(n_1,...,n_p)$
of the $p$ components $\mathbb{A}_1,...,\mathbb{A}_p$ is governed by classical MAL
\begin{equation}
  \label{eq:classMAL}
\frac{dn_{\alpha}}{dt}=\gamma_{\alpha i}\left(\overrightarrow{k}_i n_1^{\mu_{1i}}...n_p^{\mu_{pi}}-\overleftarrow{k}_i n_1^{\nu_{1i}}...n_p^{\nu_{pi}}\right),
\end{equation}
where  $\overrightarrow{k}_i, \overleftarrow{k}_i$ are the rate coefficients of the forward and the backward $i$-th reaction.

Our  objective is to explore the possibility to use the experience and the tools developed in thermodynamics (in particular in multiscale thermodynamics \cite{pkg},\cite{miroslav-MT}) to investigate (\ref{reac}). In particular, we intend to investigate the process of reducing  detail mathematical formulations of the dynamics involved in Eq. (\ref{reac}) to formulations that ignore unimportant details.

\section{Extending Mass Action Law}\label{MAL}

Thermodynamics is a theory of relations among different levels of investigation  of macroscopic systems. An upper level theory (i.e., a theory involving more details) is reduced to a lower level theory (i.e., a theory involving less details) by recognizing in the upper phase portrait a pattern that is then identified with the lower phase portrait.  The main tools in the pattern recognition process are the concept of entropy and breaking time symmetry. Boltzmann \cite{Boltzmann} has shown that the  pattern emerges by following a modified upper time evolution generated by a modified  upper vector field involving  an extra term that breaks the time symmetry and makes the  entropy (a potential of a non-mechanical origin) to increase. The pattern is revealed when the entropy reaches its maximum.

In macroscopic systems composed of particles governed by classical or quantum mechanics, the upper time evolution is driven by a gradient of energy and is time reversible. The extra term modifying it is driven by a gradient of entropy and is time irreversible.

In chemical kinetics, the MAL time evolution is driven by the gradient of entropy and is time irreversible. To bring to MAL the methods of multiscale thermodynamics, we first extend it to a theory that possesses the structure of mechanical theories investigated in multiscale thermodynamics. Having Poisson bracket for the extension, we lift the description to the kinetic theory, Liouville-like equation. As a next step, we consider two types of dissipation (Fokker-Planck-like and BGK-like) and then explore a Grad like hierarchy of the extensions of MAL theory. Hence, using the tools of multiscale thermodynamics, we reduce the kinetic theory formulation of MAL to new forms that are  suitable for investigating the qualitative properties of their solutions. The classical mass action law \eqref{eq:classMAL} will emerge as its reduced version while we will identify the effects of the added fields (reaction fluxes and correlations of fluxes and concentrations).

\subsection{The space of compositions and chemical momenta}
The space of state variables in MAL is parametrized by the numbers of moles of the individual species, $n_\alpha$. Taking inspiration from mechanics, we attach to each $n_{\alpha}$ a ``chemical momentum'' $m_{\alpha}$. The state variables are $(\nn,\mm)\in T^*\mathbb{R}^p$, where $T^*\mathbb{R}^p$ is the cotangent bundle with the base space $\mathbb{R}^p$; $\nn\in \mathbb{R}^p$. The kinematics of $T^*\mathbb{R}^p$ is expressed \cite{arnoldbook} in the canonical Poisson bracket
\begin{subequations}
\begin{equation}\label{1}
\{a,b\}=\left(\begin{array}{cc}a_{\nn}&a_{\mm}\end{array}\right)L\left(\begin{array}{cc} b_{\nn}\\b_{\mm}\end{array}\right),
\end{equation}
with
\begin{equation}
  L=\left(\begin{array}{cc}0&I\\-I&0\end{array}\right)
\end{equation}
\end{subequations}
where $a$ and $b$ are sufficiently regular real valued functions of $(\nn, \mm)$ and $I$ is the unit matrix. We use hereafter a shorthand notation: $a_{\nn}=\frac{\partial a}{\partial\nn}$ and similarly $a_{\mm}=\frac{\partial a}{\partial \mm}$. In the case when the variables on which the functions $a$ and $b$  depend are functions themselves (i.e., if they are elements of  an infinite dimensional space as it is the case in Section \ref{KMAL}),  the derivatives  are  appropriate functional derivatives  \cite{pkg}.

From the physical point of view, the Poisson bracket (or alternatively the Poisson bivector $L$) expresses mathematically the kinematics of the state variables. For comparison,  we recall that in classical mechanics the state variables are $(r,p)$ denoting the position vector and momentum of a particle. In the context of chemical kinetics, we are replacing $r$ with $\nn$ and $p$ with $\mm$.

\subsection{The space of compositions and fluxes}
Next, we reduce $(\nn,\mm) $ to the state variables (\ref{n}) because when we have a number of reactions $q$ and a number of components $p$, there are at most $p$ independent reaction steps in each reaction network involving $p$ components. The remaining ones are enslaved to these via conservation laws (as in, e.g., Michaelis-Menten kinetics). Hence, we carry out a reduction $(\nn,\mm)\mapsto (\nn,\mathcal{I})$, where
\footnote{The mapping is indeed a reduction, assuming $q \leq p$ and $rank(\gamma)=q$.}
\begin{equation}\label{2}
\left(\begin{array}{cc}\nn\\ \mathcal{I}\end{array}\right)=\left(\begin{array}{cc}I&0\\0&\gamma^T \end{array}\right)\left(\begin{array}{cc}\nn\\ \mm\end{array}\right),
\end{equation}
with $\gamma^T:\mathbb{R}^p\rightarrow \mathbb{R}^q$ being the transpose of the stoichiometric matrix (\ref{gamma}). The manifold
of the state variables
\begin{equation}\label{n}
(\nn,\mathcal{I})=(n_1,\cdots,n_p, \iota_1,...,\iota_q)
\end{equation}
is called $M$.
Note that $n_{\alpha}$ is the number of moles of the $\alpha$-th component;  $\alpha=1,\cdots,p$ and $\iota_j$ is the flux corresponding to  $j$-th reaction; $j=1,...,q$. In the classical mass action law, the fluxes $\mathcal{I}$ are not independent variables, as they are functions of $\nn$. The functions $\mathcal{I}(\nn)$, called constitutive relations, will arise in the analysis of solutions of the extended mass action law, where $\mathcal{I}$ are not functions of $\nn$, as they have their own evolution equations.

With $\mathcal{I}$  considered as an independent state variable, we are making possible to introduce   inertia and time-reversibility  into chemical kinetics, which has been observed in \cite{Cafel}. The situation is analogical to the situation in mechanics. The classical mechanics of particles includes inertia and is completely time-reversible. The experience collected in statistical mechanics shows that a very useful tool for  recognizing the overall features of particle trajectories in  systems involving very many particles interacting via very many very nonlinear forces is to brake the time symmetry. The overall pattern is then revealed by following the time evolution governed by  equations involving a new time irreversible term. We suggest that the same tool can be useful in chemical kinetics. Here the time evolution is completely time irreversible. The problem is to recognize an overall pattern of behaviour of a system consisting of  very many components making  very many very nonlinear transformations taking place on different time scales. The situation is thus opposite  to the situation in mechanics. The symmetry breaking that we are suggesting is directed  from irreversible to reversible, while in mechanics the direction is reversed. Since  inertia goes in mechanics hand in hand with the time reversibility, we introduce also inertia into chemical kinetics when we break the time symmetry. The time symmetry breaking in chemical kinetics is also discussed in \cite{GorbanTsym} but the reaction rates are not considered as independent state variables.

We now deduce the Poisson bracket expressing kinematics of $(\nn,\mathcal{I})$.
In the Poisson bracket (\ref{1}) we restrict the functions $a$ and $b$  to those that depend on $(\nn,\mm)$ only through their dependence on the moments $(\nn,\mathcal{I})$ given in (\ref{2}). This means that the derivatives in (\ref{1}) become $\frac{\partial}{\partial \nn}\rightarrow \frac{\partial}{\partial\nn}; \frac{\partial}{\partial\mm}\rightarrow \gamma\frac{\partial}{\partial \mathcal{I}}$.
With these derivatives, the Poisson bracket (\ref{1}) becomes
\begin{eqnarray}\label{pb2}
\{a,b\}&=&\left(\begin{array}{cc}a_{\nn}&a_{\mathcal{I}}\end{array}\right)
\left(\begin{array}{cc}0&\gamma\\-\gamma^T&0\end{array}\right)\left(\begin{array}{cc}b_{\nn}\\b_{\mathcal{I}}\end{array}\right)
\nonumber \\
&&=\left(\begin{array}{cc}a_{\nn}&a_{\mathcal{I}}\end{array}\right)L\left(\begin{array}{cc}b_{\nn}\\b_{\mathcal{I}}\end{array}\right).
\end{eqnarray}
Since (\ref{pb2}) was obtained just by restricting the class of functions $a$ and $b$ and  the result  involves only $(\nn,\mathcal{I})$, the bracket (\ref{pb2}) is a Poisson bracket. It can indeed be directly verified that   the following relations hold:
\begin{eqnarray}\label{propPB}
&&(i)~ \{a,b\}=-\{b,a\},\nonumber \\
&&(ii)~ \{ra+sb,c\}=r\{a,c\}+s\{b,c\}, \,\, where \,\,r\in\mathbb{R}, s\in\mathbb{R},\nonumber \\
&&(iii)~ Leibnitz \,\,identity \,\,\{a,bc\}=\{a,b\}c+\{a,c\}b,\nonumber \\
&&(iv)~ Jacobi\,\, identity \,\,\{a,\{b,c\}\}+\{b,\{c,a\}\}+\{c,\{a,b\}\}=0.
\end{eqnarray}

\subsection{GENERIC evolution with composition and fluxes}
The equations governing the Hamiltonian time evolution of $(\nn,\mathcal{I})$ that corresponds to the kinematics  (\ref{pb2}) is
\begin{equation}\label{pb1}
\dot{a}=\{a,\phi\}, \,\,\forall a.
\end{equation}
The function
\begin{equation}\label{fe}
\phi:\mathbb{R}^p\times\mathbb{R}^q\rightarrow \mathbb{R};  \,\,(\nn,\mathcal{I})
\mapsto \phi(\nn,\mathcal{I})
\end{equation}
generating the time evolution is the thermodynamic potential which is linked to energy and entropy, see below. In the isothermal case, it becomes the Helmholtz free energy.

Written explicitly, the time evolution equations (\ref{pb1}) become
\begin{equation}\label{dnwdtt}
\left(\begin{array}{cc}\frac{d\nn}{dt}\\ \frac{d\mathcal{I}}{dt}\end{array}\right)=\left(\begin{array}{cc}0&\gamma\\-\gamma^T&0\end{array}\right)
\left(\begin{array}{cc}\nn^*\\\mathcal{I}^*\end{array}\right).
\end{equation}
By the upper index $*$ we denote  conjugate variables with respect to the thermodynamic potential $\phi(\nn,\mathcal{I})$, i.e. $\nn^*=\phi_{\nn}$ and $\mathcal{I}^*=\phi_{\mathcal{I}}$.

Now we supply the Hamiltonian time evolution with dissipation:
\begin{equation}\label{dnwdt}
\left(\begin{array}{cc}\frac{d\nn}{dt}\\ \frac{d\mathcal{I}}{dt}\end{array}\right)=\left(\begin{array}{cc}0&\gamma\\-\gamma^T&0\end{array}\right)
\left(\begin{array}{cc}\nn^*\\\mathcal{I}^*\end{array}\right)-\left(\begin{array}{cc}0\\ \vartheta_{\mathcal{I}^*}\end{array}\right).
\end{equation}
The potential
\begin{equation}\label{dpT}
\vartheta(\mathcal{I}^*, \nn,\mathcal{I})
\end{equation}
introduced in the second equation in (\ref{dnwdt}) is called a dissipation potential. It is a real valued function of $(\mathcal{I}^*, \nn,\mathcal{I})$ satisfying \cite{pkg}
\begin{eqnarray}\label{disspot}
&&\vartheta(0,\nn,\mathcal{I})=0,\nonumber \\
  &&\vartheta\mbox{\textit{ as a function of $\mathcal{I}^*$ reaches its minimum at $0$ }} 
     ,\nonumber \\
&& \vartheta \mbox{\textit{ is a convex function of $\mathcal{I}^*$ in a neighborhood of $0$.}} 
\end{eqnarray}

We note that Eq.(\ref{dnwdt}) without the second term on its right hand side is time reversible in the sense that the transformation $\mathcal{I}\rightarrow-\mathcal{I}$ made in the state space compensates the inversion of time $t\rightarrow -t$. The free energy is assumed to be invariant with respect to $\mathcal{I}\rightarrow -\mathcal{I}$.

The thermodynamic potential $\phi$ is suitable for our purposes, when we require a single potential gradient in both reversible and irreversible parts \cite{pkg, pre15}
\begin{equation}
    \phi = e - T_0 s,
\end{equation}
where $e$ is the energy, $s$ entropy, and $T_0$ is the equilibrium temperature approached by relaxation of an isolated system towards the thermodynamic equilibrium.

To see this more explicitly, the GENERIC form of the evolution equations of state variables $\xx$ is \cite{go,og,hco,pkg}
\begin{equation*}
  \frac{\partial \xx}{\partial t} = L \frac{\partial e}{\partial \xx} + \frac{\delta \Xi}{\delta \frac{\partial S}{\partial \xx}},
\end{equation*}
or in terms of the thermodynamic potential gradient
\begin{equation*}
  \frac{\partial \xx}{\partial t} = T_0 L \frac{\partial \phi}{\partial  \xx} - \frac{\delta \Xi}{\delta \frac{\partial \phi}{\partial \xx}}.
\end{equation*}
This equivalence is due to the exclusive role of energy in the first term and of entropy in the second term being a consequence of the degenerative conditions of the operator $L$ and dissipation potential $\Xi$: the Poisson bracket does not affect the evolution of entropy (entropy is a Casimir of that bracket) while a dissipation potential does not change the energy, which is thus conserved.

To simplify the notation, we set $T_0=1$ hereafter. As a result, the star in the first term on the right hand side of (\ref{dnwdt}) denotes the conjugation with respect to the energy and the star in the second term on the right hand side of (\ref{dnwdt}) denotes the conjugation with respect to the entropy potential. Note that in the isothermal case, the potential becomes the free energy.



\subsection{GENERIC evolution of the classical MAL}
Hence, we have completed a formulation of MAL extension within multiscale thermodynamics. To complete the discussion we recall MAL expressed within this framework as well because it is essential for the whole study. 
A direct verification shows (see more in \cite{grchemkin}, \cite{pkg}) that
\begin{equation}\label{Phicond}
\phi(n,\mathcal{I})=\phi(n)=n_{\alpha}\ln n_{\alpha}+Q_{\alpha}n_{\alpha},
\end{equation}
and
\begin{eqnarray}\label{Thetacond}
&& \vartheta\mbox{\textit{ satisfies (\ref{disspot}) and is independent of $\mathcal{I}$}},\nonumber\\
&& \xi(\nn^*,\nn)=W_i(\nn)\left(e^{\frac{1}{2}\gamma^T\nn^*}+e^{-\frac{1}{2}\gamma^T\nn^*}-2\right),
\end{eqnarray}
where $Q_1,...,Q_p$ and $W_1,...,W_q$ are parameters (functions of $\nn $) that are related to the rate coefficients $\overrightarrow{k}_i, \overleftarrow{k}_i$ of the forward and backward $i$-th reaction by
$$\overleftarrow{k}_i=\frac{1}{2}W_i(\nn)e^{\frac{1}{2}\gamma_{\alpha i}(Q_{\alpha}+1)}\left(n_1^{\nu_{1i}}...n_p^{\nu_{p i}}n_1^{\mu_{1i}}...n_p^{\mu_{p i}}\right)^{\frac{1}{2}},$$
$$\frac{\overleftarrow{k}_i}{\overrightarrow{k}_i}=e^{\left(\gamma_{\alpha i}(Q_{\alpha} +1)\right)}.$$

With these potentials, the purely irreversible evolution of reaction kinetics
\begin{equation}\label{eq:class1}
\frac{d\nn}{dt}=-\xi_{\nn^*}(\nn^*,\nn,\mathcal{I}),
\end{equation}
becomes indeed
\begin{equation}\label{GW1}
\frac{dn_{\alpha}}{dt}=\gamma_{\alpha i}\iota^{(GW)}_i,
\end{equation}
where the Guldberg-Waage fluxes $\iota^{(GW)}_i$ are given by
\begin{equation}\label{GW2}
\iota^{(GW)}_i=\overrightarrow{k}_i n_1^{\mu_{1i}}...n_p^{\mu_{pi}}-\overleftarrow{k}_i n_1^{\nu_{1i}}...n_p^{\nu_{pi}}.
\end{equation}
Note that in Appendix \ref{clasmas} we discuss the connection between the extended and the classical MAL via the method of direct reductions.

In the vicinity of the thermodynamic equilibrium the thermodynamic affinities are close to zero \cite{dgm} and the MAL dissipation potential can be approximated by a quadratic dependence, 
\begin{equation}
    \xi(\nn^*,\nn)\approx \frac{1}{2}W_i(\nn)(\gamma^T\nn^*)^2.
\end{equation}
The evolution equations near equilibrium then become
\begin{equation}\label{GW1.lin}
    \frac{dn_{\alpha}}{dt}\approx\gamma_{\alpha i} W_i \gamma^T_{i\alpha} n^*_\alpha,
\end{equation}
and are thus linear in the thermodynamic affinities $\gamma^T_{i\alpha} n^*_\alpha$. This thermodynamically linearized chemical kinetics will be the common point with the following reductions of the Grad-like moment hierarchy.

\section{Mass-action-law Kinetic Theory}\label{KMAL}

The  mass action law  (\ref{dnwdt})  extends the classical mass action law 
\eqref{eq:classMAL},\eqref{eq:class1} but still does not provide a setting for including the influence of fluctuations. We would also like to formulate chemical kinetics in a form that is more suitable for investigating  the role of fluctuations in reductions. There are essentially two routes to take. The first one is to promote the state variables in (\ref{dnwdt}), (\ref{eq:class1}) to random variables and use stochastic mathematical formulations. The second is to lift  (\ref{dnwdt}), (\ref{eq:class1}) to a larger (typically infinite dimensional) space and use  geometry as the main tool.  We take the latter route.

\subsection{Liouville lift}

The standard first step on this route  is to replace the state variables $(\nn,\mathcal{I})$  in (\ref{dnwdt}) with a real-valued function $f(\nn,\mathcal{I})$ and the state variable  $\nn$ in (\ref{eq:class1}) with a real valued function $f(\nn)$. Their time evolution is then governed by equations obtained by making the Liouville lift of Eq.(\ref{dnwdt}) and  Eq.(\ref{eq:class1}). In other words, the time evolution of $f(\nn,\mathcal{I})$ is governed by the Liouville equation corresponding to Eq.(\ref{dnwdt}) and the time evolution of $f(\nn)$ is governed by the Liouville equation corresponding to (\ref{eq:class1}). We use the term Liouville lift in order to emphasize the original meaning of the Liouville equation introduced in \cite{Liouville},\cite{koopman},\cite{carleman} (see also Section 3.4 in \cite{pkg}). While the functions $f(\nn,\mathcal{I})$ and $f(\nn)$ can be interpreted as distribution functions and the Liouville equation can be seen as an entry to a stochastic analysis, there is an alternative (geometrical)  view of the Liouville lift.  The trajectories $(\nn(t),\mathcal{I}(t))$ and $\nn(t)$ in the  $p+q$ or $p$ dimensional spaces    are lifted to  infinite dimensional spaces with functions $f(\nn,\mathcal{I})$ and $f(\nn)$ serving as their elements in order to reveal
more clearly their geometrical features. In traditional statistical mechanics, both the stochastic and geometrical viewpoints are often combined. We shall refer to the mass action law in which $f(\nn,\mathcal{I})$ plays the role of the state variable as a mass-action-law kinetic theory (in short a MAL kinetic theory). Note that one can analogously proceed to a version in which $f(\nn)$ plays the role of the state variable, as e.g. in the seminal work of Kramers \cite{kramers}.

The next step in the development of the  MAL kinetic theory is to lift the time evolution equation (\ref{dnwdt}) to an equation governing the time evolution of $f(\nn,\mathcal{I})$. We begin with the Hamiltonian part of the time evolution. The Liouville equation corresponding to (\ref{dnwdt}) without the second term on its right hand side,
\begin{equation}\label{Liu1}
\frac{\partial f}{\partial t}=-\frac{\partial}{\partial n_{\alpha}}\left(f\gamma_{\alpha j}\frac{\partial \phi}{\partial \iota_j}\right)+\frac{\partial}{\partial \iota_j}\left(f\gamma_{j\alpha}^T\frac{\partial \phi}{\partial n_{\alpha}}\right),
\end{equation}
can be cast (as can be directly verified) into the form (\ref{pb1}) with
\begin{equation}\label{Lnw}
\{A,B\}=\int d\nn\int d\mathcal{I}f\left(\frac{\partial A_f}{\partial n_{\alpha}}\gamma_{\alpha j}\frac{\partial B_f}{\partial \iota_j}-\frac{\partial B_f}{\partial n_{\alpha}}\gamma_{j\alpha}^T\frac{\partial A_f}{\partial \iota_j}\right),
\end{equation}
see e.g. \cite{pkg} for details of the calculation,
replacing $\{a,b\}$  (the functions
$A$ and $B$ appearing in (\ref{Lnw}) are sufficiently regular real-valued functions of $f(\nn,\mathcal{I})$)    and with
\begin{equation}\label{phL}
\Phi(f)=\int d\nn\int d\mathcal{I} f(\nn,\mathcal{I})\phi(\nn,\mathcal{I}),
\end{equation}
replacing $\phi$.
The bracket (\ref{Lnw})  is the  Liouville lift of the Poisson bracket (\ref{pb2}). Note that the correspondence between the Liouville equation \eqref{Liu1} and the being lifted evolution equations \eqref{dnwdt} is confirmed and further discussed below in Section \ref{2nd}.

From the mathematical perspective, the Poisson bracket \eqref{pb2} represents a Lie bracket on the Hamiltonian vector fields on $M$, see, e.g., Eq. 47 in \cite{momentum-Euler}. The space of distribution functions can be seen as an element of the dual of the Lie algebra of those vector fields, see, e.g., Eq. 49 in \cite{momentum-Euler}. Therefore, the derivatives of functionals of $f$, which are functions on $M$ themselves, can be seen as elements of the Lie algebra dual. Bracket \eqref{Lnw} is then a Lie-Poisson bracket on the Lie algebra dual \cite{arnold}.

\subsection{Fokker-Planck-like dissipation}

Now we proceed to the dissipative part of the Liouville lift of the time evolution (\ref{dnwdt}). We note that if we add to the right hand side of (\ref{Liu1}) a term $-\Theta(f^*)$ where $f^*=\Phi_f$ and $\Theta$ satisfies all the properties listed in (\ref{disspot}) (except that $\vartheta$ is replaced by $\Theta$ and  $\mathcal{I}^*$ is  replaced by $f^*$ and $(\nn,\mathcal{I})$  by $f$ ) then the time evolution governed by such modified equation will imply $\dot{\Phi}\leq 0$. For the reasons  explained below in this section, we
 introduce
\begin{equation}\label{mcL}
\mathfrak{J}=(\mathcal{J}_1,...,\mathcal{J}_q)=\left(\frac{\partial f^*}{\partial \iota_1},...,\frac{\partial f^*}{\partial \iota_q}\right)=\frac{\partial f^*}{\partial\mathcal{I}},
\end{equation}
and let the dissipation potential depend on $f^*$ through its dependence on  $\mathcal{J}$. In such case $\Theta_{f^*}=-\frac{\partial\Theta_{\mathcal{J}}}{\partial \mathcal{I}}$.

The thermodynamic potential $\phi$ appearing in \eqref{Liu1} can be more general and so can the potential $\Phi$. For example, $\phi$ can be a function of other variables than just $\nn$ and $\mathcal{I}$ as we shall see below. To this end, we keep the potential $\Phi$ and the dissipation potential $\Theta$ undetermined except that they are required to satisfy the general requirements (i.e. $\Phi$ is convex and $\Theta$ satisfies (\ref{disspot})).
In particular, the thermodynamic potential $\Phi$ does not have to be the potential (\ref{phL}).
The kinetic equation governing the time evolution in the  MAL kinetic theory is thus
\begin{eqnarray}\label{Liu11}
\frac{\partial f}{\partial t}&=&-\frac{\partial}{\partial n_{\alpha}}\left(f\gamma_{\alpha j}\frac{\partial \Phi_f}{\partial \iota_j}\right)+\frac{\partial}{\partial \iota_j}\left(f\gamma_{j\alpha}^T\frac{\partial \Phi_f}{\partial n_{\alpha}}\right)
                                 +\frac{\partial\Theta_{\mathcal{J}_i}}{\partial \iota_i}\nonumber\\
                                 &=&-\frac{\partial}{\partial n_{\alpha}}\left(f\gamma_{\alpha j}\frac{\partial f^*}{\partial \iota_j}\right)+\frac{\partial}{\partial \iota_j}\left(f\gamma_{j\alpha}^T\frac{\partial f^*}{\partial n_{\alpha}}\right)
+\frac{\partial\Theta_{\mathcal{J}_i}}{\partial \iota_i},
\end{eqnarray}
and one can confirm that the corresponding Poisson bracket is still \eqref{Lnw}.

A way to introduce the dissipative terms in the equation for $f$ is to first introduce noise, e.g., via the derivative of a Wiener process, into the equations \eqref{dnwdt} and subsequently reconstruct Eq. \eqref{Liu11} as the corresponding Fokker-Planck equation, see e.g. \cite{hco}, which we discuss later in this section.

Due to the antisymmetry of the Poisson bracket, the  thermodynamic potential $\Phi$ remains unchanged during the Hamiltonian time evolution governed by (\ref{Liu1}). Other potentials that remain unchanged are Casimirs of the Poisson bracket (\ref{Lnw}) (i.e. potentials $C$ for which $\{A,C\}=0, \forall A$).  We directly verify that $\int d\nn\int d\mathcal{I} f$  is one  such  Casimir. The conservation of   $\int d\nn\int d\mathcal{I} f$  is important since it makes it possible to interpret $f(\nn,\mathcal{I})$ as a distribution function. Indeed, the  normalization of $f$ (i.e. $\int d\nn\int d\mathcal{I}f(\nn,\mathcal{I})$ ) that, in the case we interpret $f$ as a distribution function,  expresses the  sum of all probabilities, must remain constant. In order to keep the conservation of $f(\nn,\mathcal{I})$ also in the time evolution governed by (\ref{Liu11}), we let the dissipation potential depend on $f^*$ only through its dependence on $\mathfrak{J}$ introduced in (\ref{mcL}).

The most important property of solutions to (\ref{Liu11}) is (compare with (\ref{pb5}) in Appendix)
\begin{multline}\label{PH1}
  \dot{\Phi}= \int d\nn\int d\mathcal{I}f^* \frac{\partial f}{\partial t} =\int d\nn\int d\mathcal{I} \frac{\partial f^*}{\partial n_\alpha} f \gamma_{\alpha j} \frac{\partial f^*}{\partial \mathcal{I}_j} -\frac{\partial f^*}{\partial \mathcal{I}_j} f \gamma_{j\alpha}^T \frac{\partial f^*}{\partial n_\alpha}-f^* \frac{\partial \Theta_{\mathcal{J}_i}}{\partial \iota_i} \\
  =\int d\nn\int d\mathcal{I}f^*\frac{\partial \Theta_{\mathcal{J}_i}}{\partial \iota_i}=-\int d\nn\int d\mathcal{I} \mathcal{J}_i\Theta_{\mathcal{J}_i} \leq 0.
\end{multline}
This inequality, together with the convexity of $\Phi$,  makes it possible to consider  $\Phi$ as a Lyapunov function corresponding to the approach, as $t\rightarrow\infty$, of solutions to (\ref{Liu11})  to  states at which $\mathfrak{J}=0$.

Another key property that was already mentioned follows from the subsequent calculation
\begin{multline}\label{prob}
  \frac{\partial \int d\nn\int d\mathcal{I}f(\nn,\mathcal{I})}{\partial t}=  \int d\nn\int d\mathcal{I}\frac{\partial f(\nn,\mathcal{I})}{\partial t}=\\= \int d\nn\int d\mathcal{I} \Theta_{f^*} = \int d\nn\int d\mathcal{I} \frac{\partial \Theta_{\mathcal{J}_i}}{\partial \iota_i}= 0,
\end{multline}
which allows to interpret $f(\nn,\mathcal{I})$ as a distribution function.


Let us now formulate the fluctuations, noise and the Fokker-Planck equation, explicitly. Equations for $\nn$ and $\mathcal{I}$ can be equipped with a Wiener noise, see also \cite{Cafel},
\begin{subequations}\label{eq.nI.noise}
    \begin{align}
        \dot{n}_\alpha &= \gamma_{\alpha i}\frac{\partial e}{\partial \iota_i}\\
        \dot{\iota}_i &= -\gamma_{i \alpha}\frac{\partial e}{\partial n_\alpha} + M_{ij} \frac{\partial s}{\partial \iota_i} + B_{ij} \dot{W}_j,
    \end{align}
    where $e(\nn,\mathcal{I})$ is the energy of the system with composition $\nn$ and rates $\mathcal{I}$, $W_i$ is the Wiener process in the direction of the rate $\iota_i$, $M_{ij}$ is the dissipative matrix (Hessian of a quadratic dissipation potential), $s$ is the entropy, derivative of which reads $s_{\iota_i} = - e_{\iota_i}/T$ with temperature $T=e_s$, and $B_{ij}$ is the strength of the noise. The fluctuation-dissipation theorem tells, moreover, that $2k_B \mathbf{M} = \mathbf{B}\mathbf{B}^T$, see \cite{hco}.
\end{subequations}
For simplicity, we choose $B_{ij} = \delta_{ij} \sqrt{2k_B T\xi}$, which gives $M_{ij} = \delta_{ij} T\xi$ and $\xi$ then represents a friction constant.

Apart from the classical chemical energy, energy $e$ now also depends on the rates $\mathcal{I}$. A physical reasoning can be found for instance in the inertia of polarization \cite{Cafel}. Let us assume that the part of the energy dependent on the rates is the sum of contributions of the individual rates and that the contributions have the form of $\cosh$, which can be approximated by a quadratic dependence in the region of small fluxes,
\begin{equation}\label{eq.e}
    e = e^{chem}(\nn) + \sum_i \lambda_i (\cosh(K_i \iota_i)-1)
    \approx e^{chem}(\nn) + \frac{1}{2}\sum_i \lambda_i(K_i \iota_i)^2.
\end{equation}
The positive constants $\lambda_i$ and $K_i$ represent the inertia of the fluxes. This choice will become clearer later, when reducing the dynamics to the usual MAL.


The corresponding Fokker-Planck equation, the reversible part of which is generated by the Poisson bracket, reads
\begin{align}\label{eq.FP}
    \partial_t f &= -\sum_{\alpha,i}\frac{\partial}{\partial n_\alpha}\left(f \Gamma_{\alpha i}\frac{\partial e}{\partial \iota_i}\right)
     +\sum_{\alpha,i}\frac{\partial}{\partial \iota_i}\left(f \Gamma_{i \alpha}\frac{\partial e}{\partial n_\alpha}\right) \nonumber\\
     &+ \sum_{i}\frac{\partial}{\partial \iota_i}\left(f \xi \frac{\partial e}{\partial \iota_i}\right)
     +\sum_{i}\frac{\partial^2}{\partial \iota_i \partial \iota_i}\left(2 k_B T \xi f\right)
\end{align}
where the last term is a consequence of the noise term in Eqs. \eqref{eq.nI.noise}, see Appendix \ref{sec.Ito} for the details.

Alternatively, the Fokker-Planck equation \eqref{eq.FP} can be derived by a Liouville lift from the deterministic part of Eqs. \eqref{eq.nI.noise}, as described in \cite{pkg},
\begin{subequations}\label{eq.Genlift}
    \begin{align}
        E^{(f)} &= \int d\nn \int d\mathcal{I} e^{mic} f\\
        S^{(f)} &= \int d\nn \int d\mathcal{I} s f - k_B \int d\nn \int d\mathcal{I} f\ln f\\
        \{A(f),B(f)\}^{(f)} &= \int d\nn \int d\mathcal{I} f \left\{\frac{\delta A}{\delta f}, \frac{\delta B}{\delta f}\right\}\\
        \Xi^{(f)} &= \int d\nn \int d\mathcal{I} f \Xi\left(\frac{\delta f^*}{\delta (\nn, \mathcal{I})}\right),
    \end{align}
    where the Poisson bracket $\{\bullet,\bullet\}$ is given in \eqref{pb2} and the dissipation potential is $\Xi = \frac{1}{2}T\xi (\II^*)^2$.
\end{subequations}
The Fokker-Planck equation is then the GENERIC evolution implied by the above building blocks.

Another option how to ensure the approach of the distribution function to the equilibrium value is the Hamiltonian dynamics combined with and BGK dynamics, which we discuss in Appendix \ref{BGKdis}.

\section{Grad-like hierarchy of the  MAL kinetic theory }\label{2nd}

\subsection{Grad-like moments}
In this section, we make an observation indicating a possibility to reduce  the MAL kinetic theory to a finite dimensional theory that still takes into account fluctuations. First we construct a hierarchy of moments, 
\begin{subequations}
    \begin{align}
                                        &\int d\nn\int d \mathcal{I} f(\nn,\mathcal{I}) n_{\alpha_1} n_{\alpha_2} \dots \\
                                                              &\int d\nn\int d \mathcal{I} f(\nn,\mathcal{I}) I_{i_1} I_{i_2} \dots .
    \end{align}
These latter kind of moments are the analogy of the moments taken in the Grad hierarchy \cite{Grad}, i.e. with respect to the momentum, while the former kind of moments would be analogical to the (usually not considered) moments with respect to the spatial coordinate.
\end{subequations}

Let us, in particular, analyze the hierarchy restricted to the second moments, containing actually $1+3+6+9+6=25$ fields. Poisson bracket (\ref{Lnw}) can be exactly reduced to a Poisson bracket expressing the kinematics of the following moments of $f(\nn,\mathcal{I})$:
\begin{subequations}\label{moments}
\begin{eqnarray}
\mu&=&\int d\nn\int d\mathcal{I}f(\nn,\mathcal{I})\\
\nu_\alpha&=& \int d\nn\int d\mathcal{I} n_\alpha f(\nn,\mathcal{I}) \\
\zeta_i&=&\int d\nn\int d\mathcal{I} \iota_i f(\nn,\mathcal{I}) \\
a_{ij}&=&\int d\nn\int d\mathcal{I}\iota_i \iota_j f(\nn,\mathcal{I}) \\
b_{\alpha \beta}&=&\int d\nn\int d\mathcal{I}n_{\alpha} n_{\beta} f(\nn,\mathcal{I}) \\
c_{\alpha j}&=&\int d\nn\int d\mathcal{I}n_{\alpha} \iota_j f(\nn,\mathcal{I})\\
    s &=& \int d\nn \int d \mathcal{I} \eta(f),
\end{eqnarray}
    including entropy ($\eta(f)$ being a smooth real-valued function). These moments resemble the Grad hierarchy and $\mu$ represents the average mass\footnote{Note that the stoichiometric coefficients also are denoted by $\mu_\alpha$, i.e. with indexes.}, $\nu_\alpha$ number of moles, $\zeta_i$ the rate of the $i-$th reaction, $a_{ij}$ correlations of the rates, $b_{\alpha \beta}$ correlations of the compositions, and $c_{\alpha j}$ correlations between the rates and compositions.
\end{subequations}

\subsection{Hamiltonian evolution of the moments}
By replacing $\frac{\partial}{\partial f(\nn,\mathcal{I})}$ in  (\ref{Lnw}) with
$$\frac{\partial}{\partial f(\nn,\mathcal{I})}\rightarrow \frac{\partial}{\partial \mu}+n_{\alpha}\frac{\partial}{\partial \nu_{\alpha}}+\iota_j\frac{\partial}{\partial \zeta_j}
+n_{\alpha}n_{\beta}\frac{\partial}{\partial b_{\alpha \beta}}+\iota_i\iota_j\frac{\partial}{\partial a_{ij}}+n_{\alpha} \iota_i\frac{\partial}{\partial c_{\alpha i}}$$
we arrive at
\begin{eqnarray}\label{BBB}
\{A,B\}&=&\mu\gamma_{\alpha j}\left[A_{\nu_{\alpha}}B_{\zeta_j}-B_{\nu_{\alpha}}A_{\zeta_j}\right]\nonumber \\
&&+2\zeta_k \gamma_{\alpha j}\left[A_{\nu_{\alpha}}B_{a_{kj}}-B_{\nu_{\alpha}}A_{a_{kj}}\right]\nonumber \\
&&+\nu_{\beta}\gamma_{\alpha j}\left[A_{\nu_{\alpha}}B_{c_{\beta j}}-B_{\nu_{\alpha}}A_{c_{\beta j}}\right]\nonumber \\
&&+2\nu_{\beta}\gamma_{\alpha j}\left[A_{b_{\alpha \beta}}B_{\zeta_j}-B_{b_{\alpha \beta}}A_{\zeta_j}\right]\nonumber \\
&&+4c_{\beta k}\gamma_{\alpha j}\left[A_{b_{\alpha\beta}}B_{a_{kj}}-B_{b_{\alpha\beta}}A_{a_{kj}}\right]\nonumber \\
&&+2b_{\beta\epsilon}\gamma_{\alpha j}\left[A_{b_{\alpha\beta}}B_{c_{\epsilon j}}-B_{b_{\alpha\beta}}A_{c_{\epsilon j}}\right]\nonumber \\
&&+\zeta_i\gamma_{\alpha j}\left[A_{c_{\alpha i}}B_{\zeta_j}-B_{c_{\alpha i}}A_{\zeta_j}\right]\nonumber \\
&&+2a_{ik}\gamma_{\alpha j}\left[A_{c_{\alpha i}}B_{a_{kj}}-B_{c_{\alpha i}}A_{a_{kj}}\right]\nonumber \\
&&+c_{\epsilon i}\gamma_{\alpha j}\left[A_{c_{\alpha i}}B_{c_{\epsilon j}}-B_{c_{\alpha i}}A_{c_{\epsilon j}}\right]
\end{eqnarray}
which is a bracket involving only the moments (\ref{moments}). This means that if we restrict the functions $A$ and $B$ in (\ref{Lnw}) to those that depend on $f(\nn,\mathcal{I})$ only through their  dependence on the moments (\ref{moments}) then we arrive at the bracket  (\ref{BBB}) that involves only the moments (\ref{moments}). Consequently, the fact that (\ref{Lnw}) is a Poisson bracket implies that the bracket (\ref{BBB}) is also a Poisson bracket.

The Hamiltonian time evolution equations corresponding to this Poisson bracket are
\begin{subequations}\label{ncba}
\begin{eqnarray}
\frac{d\mu}{dt}&=&0, \\
\frac{d\nu_{\alpha}}{dt}&=&\gamma_{\alpha j}\mu e_{\zeta_j}\\
&&+\nu_{\beta}\gamma_{\alpha j} e_{c_{\beta j}}+2\zeta_k\gamma_{\alpha j} e_{a_{kj}},\\
\frac{d\zeta_j}{dt}&=&-\gamma_{j\alpha}^T\mu e_{\nu_{\alpha}}\\
&&-\zeta_i\gamma_{j\alpha}^T e_{c_{\alpha i}}   -2\nu_{\beta}\gamma_{j\alpha}^T e_{b_{\alpha\beta}},\\
\frac{db_{\alpha\beta}}{dt}&=& 2\nu_{\beta}\gamma_{\alpha j} e_{\zeta_j}+4c_{\beta k}\gamma_{j\alpha}^T e_{a_{kj}}+2b_{\beta\epsilon}\gamma_{\alpha j}e_{c_{\epsilon j}},\\
\frac{d a_{kj}}{dt}&=&-2\zeta_k\gamma_{j\alpha}^T e_{\nu_{\alpha}}-4c_{\beta k}\gamma_{j\alpha}^T e_{b_{\alpha\beta}}-2a_{ik}\gamma_{j\alpha}^T e_{c_{\alpha i}},\\
\frac{d c_{\beta j}}{dt}&=& -\nu_{\beta}\gamma_{j\alpha}^T e_{\nu_{\alpha}}+\zeta_j\gamma_{\beta i} e_{\zeta_i} -c_{\beta i}\gamma_{j\alpha}^T e_{c_{\alpha i}}+c_{\epsilon j}\gamma_{\beta i} e_{c_{\epsilon i}} \\
&&-2b_{\epsilon \beta}\gamma_{j\alpha}^T  e_{b_{\alpha \epsilon}}+2a_{jk}\gamma_{\beta i} e_{a_{ki}},\\
    \frac{d s}{d t} &=& 0,
\end{eqnarray}
where $e(\mu,\nu_\alpha,\zeta_j,b_{\alpha\beta},a_{kj},c_{\beta j},s)$ is the energy expressed in terms of the moments. Note that the temperature is given by the derivative with respect to entropy, $T= e_s$.
\end{subequations}

Note that the zeroth moment $\mu$ is conserved as we investigated above ($f$ is a distribution function). Furthermore, the energy is conserved automatically due to the antisymmetry of the bracket, and that the entropy is conserved as well, being a Casimir of that bracket.
If one chooses only the zeroth and first moments, i.e., $\mu, \nu,\zeta$, one obtains the kinetic theory analogue of the classical extended MAL evolution \eqref{dnwdt}. We see indeed that the second and the third equation with only the first terms on their right hand sides are exactly the governing equations  (\ref{dnwdt}) of the  MAL theory. The remaining terms on their right hand sides  represent the influence of fluctuations.

\subsection{Gaussian noise and the Fokker-Planck-like dissipation}

Let us now turn to the overall evolution of the moments \eqref{moments}.
The reversible part of the evolution of the moments, Eqs. \eqref{ncba}, is not affected by the dissipative terms (the latter two in Eq. \eqref{eq.FP}). However, the dissipative terms add an irreversible evolution of the moments, which is given by the projection of the irreversible terms to those moments. 

\subsubsection{First moments}
Let us for the moment focus on the moments $\mu$, $\nnu$, and $\zzeta$, taking energy independent of the higher moments. Moreover, we shall get restricted to the isothermal processes and so we do not need the equation for entropy, which can be evaluated based on the temperature and other state variables.
The overall equations for the first moments are then the sum of the reversible and irreversible contributions,
\begin{subequations}\label{eq.moments.first}
\begin{eqnarray}
\frac{d\mu}{dt}&=&0 \\
\frac{d\nu_{\alpha}}{dt}&=&\gamma_{\alpha j}\mu e_{\zeta_j}\\
    \frac{d\zeta_j}{dt}&=&-\gamma_{j\alpha}\mu e_{\nu_{\alpha}}  - \xi \lambda_j K^2_j \zeta_j.
\end{eqnarray}
\end{subequations}
Note that the irreversible contributions are yet to be made explicit for instance by the Chapman-Enskog procedure \cite{dgm,chapman} or by another approximation \cite{struch}.

\begin{figure}
    \begin{center}
        \includegraphics[scale=0.5]{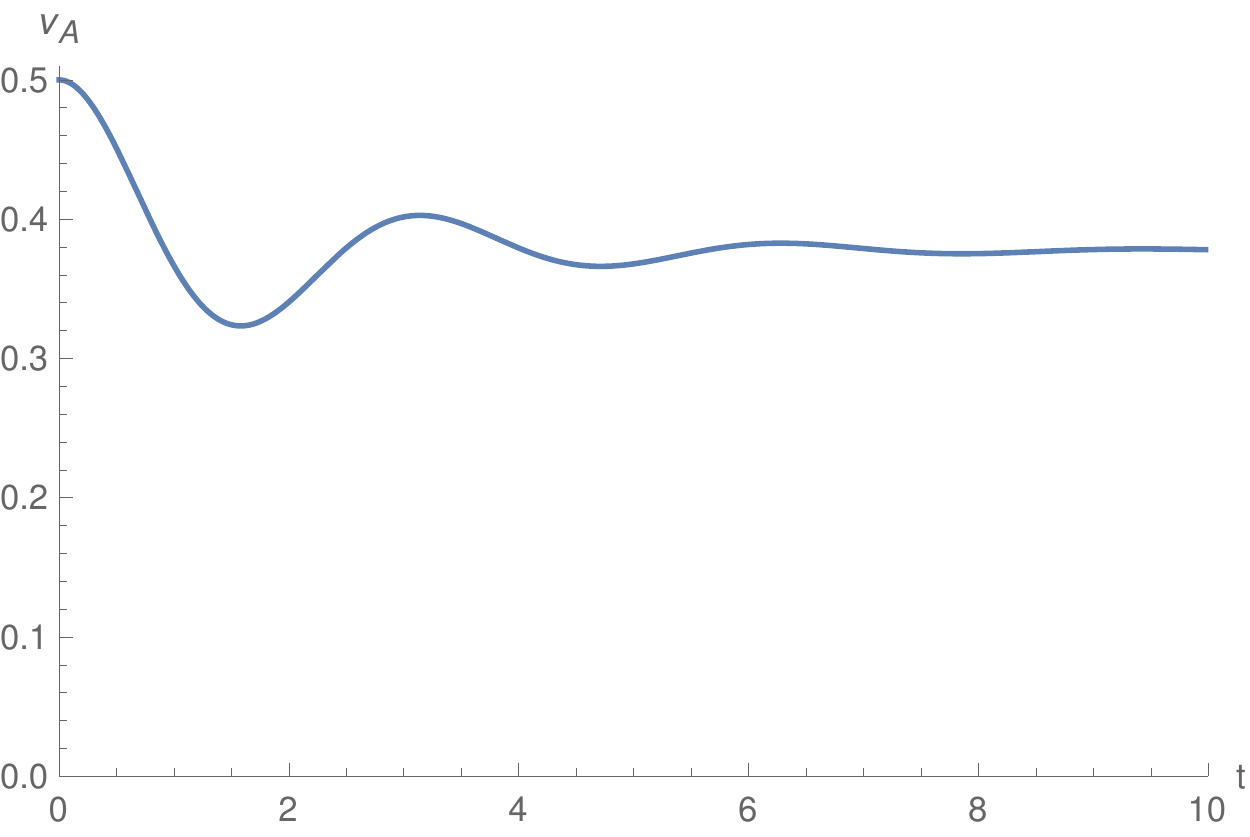}
        \caption{\label{fig.oscillations}A typical behavior of Eqs. \eqref{eq.moments.first}, showing damped oscillations. In the case of strong dissipation, however, the oscillations disappear and we get a purely monotonous behavior as in the classical MAL. Parameters in this figure were chosen for a prototypical reaction $A\leftrightarrow B$, $e = \nu_A + \nu_A\ln(1-\nu_A) + 0.5 (1-\nu_A) + (1-\nu_A)\ln(1-\nu_A)$, $K_j = \lambda_j =\xi=1$. See \cite{pkg} for a derivation of such formulas for the chemical energy.}
    \end{center}
\end{figure}

See Fig. \ref{fig.oscillations} for a demonstration of the inertial feature of Eqs. \eqref{eq.moments.first}.
One can clearly see the possibility for damped chemical oscillations, exhibiting inertial effects.

\subsubsection{Second moments}
Similarly, the overall equations for the first and second moments are then the sum of the reversible and irreversible contributions. For simplicity we shall restrict the energy to the quadratic approximation in the second moments \eqref{eq.e},
\begin{subequations}\label{eq.moments.dis}
\begin{eqnarray}
\frac{d\mu}{dt}&=&0 \\
\frac{d\nu_{\alpha}}{dt}&=&\gamma_{\alpha j}\mu e_{\zeta_j}
+\nu_{\beta}\gamma_{\alpha j}e_{c_{\beta j}}+2\zeta_k\gamma_{\alpha j}e_{a_{kj}} \\
\frac{d\zeta_j}{dt}&=&-\gamma_{j\alpha}\mu e_{\nu_{\alpha}}
-\zeta_i\gamma_{j\alpha}e_{c_{\alpha i}}   -2\nu_{\beta}\gamma_{j\alpha}e_{b_{\alpha\beta}}
       -\frac{\xi}{2m_j} \zeta_j \\
\frac{db_{\alpha\beta}}{dt}&=& 2\nu_{\beta}\gamma_{\alpha j}e_{\zeta_j}+4c_{\beta k}\gamma_{j\alpha}e_{a_{kj}}+2b_{\beta\epsilon}\gamma_{\alpha j}e_{c_{\epsilon j}}\\
\frac{d a_{kj}}{dt}&=&-2\zeta_k\gamma_{j\alpha}e_{\nu_{\alpha}}-4c_{\beta k}\gamma_{j\alpha}e_{b_{\alpha\beta}}-2a_{ik}\gamma_{j\alpha}e_{c_{\alpha i}}\nonumber\\
    &&-\xi\left(\frac{1}{m_k}+\frac{1}{m_j}\right)a_{kj}+\delta_{kj}2k_B T\xi \mu\\
\frac{d c_{\beta j}}{dt}&=& -\nu_{\beta}\gamma_{j\alpha}e_{\nu_{\alpha}}+\zeta_j\gamma_{\beta j}e_{\zeta_i} -c_{\beta i}\gamma_{j\alpha}e_{c_{\alpha i}}+c_{\epsilon j}\gamma_{\beta i}e_{c_{\epsilon i}} \nonumber\\
&&-2b_{\epsilon \beta}\gamma_{j\alpha} e_{b_{\alpha \epsilon}}+2a_{jk}\gamma_{\beta i}e_{a_{ki}}
    -\frac{\xi}{m_j}c_{\beta j},
\end{eqnarray}
where we no longer write the equation for entropy, assuming constant temperature. The effective masses are defined as $\frac{1}{2m_j} =\lambda_j K^2_j$, see \cite{Cafel} for a relation between the effective masses and polarization dynamics. Note that $\bar{m}^{-1} = \sum_i \frac{1}{m_i}$.
\end{subequations}
An explicit expression for the entropy production can be obtained using the principle of maximum entropy, as well as an explicit expression for the energy $e$ dependent on the moments. However, we shall leave this for future research, since if we explictly consider variations of temperature, the noise becomes non-constant, and we would have to choose an interpretation for the stochastic differential equation (Ito, Stratonovich, or Klimontovich) \cite{Klimontovich1990}.

\subsubsection{Connection to classical MAL. Reductions}


Let us now discuss connection to the classical MAL via reductions. We start by showing that the governing equations for the first moments can be reduced exactly to MAL.
Using asymptotic expansion 
\begin{equation}\label{eq.exp}
    \zeta_j = \zeta_{j,0}\left(\zeta^{(0)}_j + \delta \zeta^{(1)}_j + \dots\right)
\end{equation}
in the small parameter $\delta_j = \frac{\zeta_{j,0}^2}{\lambda_j T_0 \xi} \ll 1$, where $T_0$ is the characteristic time-scale and $\zeta_{j,0}$ is the characteristic magnitude of the reaction flux $\zeta_j$. 
The equation for $\zeta_j$ then becomes 
\begin{align}
    \dot{\zeta}^{(0)}_j  +\delta^1 \dot{\zeta}^{(1)}_j + \dots = -\gamma_{j\alpha}\mu T_0 e_{\nu_{\alpha}} - \delta^{-1} K_j\zeta_{j,0}^2 \left(K_j \zeta_{j,0}(\zeta^{(0)}_j + \delta^1\zeta^{(1)}_j)\right)+\dots,
\end{align}
and by comparing the terms of order $\delta^{-1}$ and $\delta^{0}$ we obtain
\begin{equation}
    \zeta^{(0)}_j = 0
    \quad\mbox{and}\quad
    \zeta^{(1)}_j = -\frac{T_0}{\zeta_{j,0}^2 K^2_j}\sum_\alpha \gamma_{j\alpha}\mu e_{\nu_{\alpha}}.
\end{equation}
The asymptotic rate is then 
\begin{equation}
    \zeta_j \approx  -\frac{1}{\xi \lambda_j K^2_j}\sum_\alpha \gamma_{j\alpha}\mu e_{\nu_{\alpha}}.
\end{equation}
Plugging this back into the equation for $\nu_\alpha$ we get
\begin{equation} \label{MALfromKinTheo}
\frac{d\nu_{\alpha}}{dt} = \sum_j \gamma_{\alpha j}\mu \lambda_j K_j \zeta_j
\approx
    \sum_j \gamma_{\alpha j}\mu \lambda_j K_j \left(-\frac{\lambda_j}{\xi K^2_j}\sum_\alpha \gamma_{j\alpha}\mu e_{\nu_{\alpha}}\right),
\end{equation}
which is a form of the thermodynamically linearized MAL \eqref{GW1.lin}.

\paragraph{Second moments.}

If we use the Dynamic Maximum Entropy method \cite{DynMaxEnt} in the conjugate or direct variables for the identification of the reduced evolution while assuming the energy to be quadratic in the second moments, we get a decoupled evolution of the second moments from the lower moments.  Then the evolution equations for the second moments provide a relation for their conjugates in terms of the lower level state variables (zeroth and first moments) and their conjugates, $e_{a_{ki}}(\nu,e_{\nu},\zeta,e_{\zeta})$, $e_{b_{\alpha\beta}}(\nu,e_{\nu},\zeta,e_{\zeta})$, and $e_{c_{\beta j}}(\nu,e_{\nu},\zeta,e_{\zeta})$.

In particular, with the reduction in direct variables we get  $a_{ki}=b_{\alpha\beta}=c_{\beta j}=0$ from MaxEnt. A more reasonable estimate (not requiring all the correlations to disappear) is obtained by MaxEnt in the conjugate variables where due to the quadratic dependence in energy we obtain $e_{a_{ki}}=e_{b_{\alpha\beta}}=e_{c_{\beta j}}=0$ while the direct variables follow (decoupled) evolution equations
\begin{eqnarray*}
\frac{db_{\alpha\beta}}{dt}&=& 2\nu_{\beta}\gamma_{\alpha j}e_{\zeta_j}\\
\frac{d a_{kj}}{dt}&=&-2\zeta_k\gamma_{j\alpha}e_{\nu_{\alpha}}\\
\frac{d c_{\beta j}}{dt}&=& -\nu_{\beta}\gamma_{j\alpha}e_{\nu_{\alpha}}+\zeta_j\gamma_{\beta j}e_{\zeta_i},
\end{eqnarray*}
and the first moments are governed by
\begin{eqnarray*}
\frac{d\mu}{dt}&=&0 \\
\frac{d\nu_{\alpha}}{dt}&=&\gamma_{\alpha j}\mu e_{\zeta_j}\\
    \frac{d\zeta_j}{dt}&=&-\gamma_{j\alpha}\mu e_{\nu_{\alpha}}   -\frac{\xi}{2m_j} \zeta_j.
\end{eqnarray*}
The governing equations for the zeroth and first moments are exactly the extended MAL equations \eqref{dnwdt} which were shown above to reduce to the classical MAL. Hence the Dynamic Maximum Entropy method yields a connection between the Grad-like hierarchy of chemical kinetics and the classical and extended MAL.

In the remaining text we use asymptotic analysis to reveal subtle corrections stemming from the second moments which were not revealed by the Dynamic Maximum Entropy Method. We again perform the asymptotic analysis by expanding the second moments as in Eq. \eqref{eq.exp}. Again, the presence of a single $\delta^{-1}\propto \xi$ term in the evolution equations for $c_{\beta j}$ and $\zeta_j$ results in the observation that the zeroth order solution is zero and that the leading order contribution is constant in time, i.e. a quasi steady state is reached in these state variables while its value is proportional to the asymptotic parameter $\delta=\frac{\overline{m}}{T_0 \xi}$. Note that indeed it is required that the equilibrium value of these state variables is zero, since they are odd with respect to the time reversal transformation. For example, for $c_{\beta j}$ holds
\begin{multline*}
  c_{\beta j}\approx \frac{m_j}{\xi}\big(-\nu_{\beta}\gamma_{j\alpha}e_{\nu_{\alpha}}+\zeta_j\gamma_{\beta j}e_{\zeta_i} \\-c_{\beta i}\gamma_{j\alpha}e_{c_{\alpha i}}+c_{\epsilon j}\gamma_{\beta i}e_{c_{\epsilon i}}-2b_{\epsilon \beta}\gamma_{j\alpha} e_{b_{\alpha \epsilon}}+2a_{jk}\gamma_{\beta i}e_{a_{ki}}\big)=O(\delta).
\end{multline*}
The equation for the even state variable $a_{kj}$ contains two $O(\delta^{-1})$ terms allowing for a dominant balance yielding
\begin{equation*}
  a_{kj}^{(1)}= 2 \delta_{kj} k_B T \mu \frac{m_k m_j}{m_j+m_k}=O(1).
\end{equation*}
The remaining time evolutions of $\nu_\alpha$ and $b_{\alpha \beta}$ is $O(\delta)$ and hence they are approximately constant (but nonzero) at the leading order at the chosen time scale $T_0$.




Finally, we would like to identify the corrections to the thermodynamically linearized MAL dynamics explicitly. We already know that the relaxation of reaction fluxes $\zeta_j$ (first moments) provides one such correction (while reaching MAL with $\zeta_j$ vanishing as shown in the previous subsection). 
With the knowledge of the leading order behaviour of the second moments, we can estimate the evolution and correction to the MAL due to the correlations as follows:
\begin{eqnarray*}
  \frac{d\zeta_j}{dt}&=&-\gamma_{j\alpha}\mu e_{\nu_{\alpha}} -\underbrace{\zeta_i}_{O(\delta)}\gamma_{j\alpha}\underbrace{e_{c_{\alpha i}}}_{O(\delta)}   -2\nu_{\beta}\gamma_{j\alpha} e_{b_{\alpha\beta}} -\underbrace{\xi \lambda_j K^2_j}_{O(\delta^{-1})} \underbrace{\zeta_j}_{O(\delta)}  \\
                     &\approx& -\gamma_{j\alpha}\mu e_{\nu_{\alpha}} -2\nu_{\beta}\gamma_{j\alpha}e_{b_{\alpha\beta}} -\xi \lambda_j K^2_j \zeta_j.
\end{eqnarray*}
Hence the correlations affect the evolution of the reaction fluxes $\zeta_j$ and this correction via $e_{b_{\alpha \beta}}$ is subleading provided the slowly varying (on the $\delta$ time scale) typical values of $b_{\alpha \beta}$ are smaller than the magnitude of the total mass $\mu$. Note that this implies that one can expect the correlations to be more significant in sparser and diluted systems where the law of large numbers does not apply. Repeating the reduction of the first moment $\zeta$ to the classical MAL while retaining the correction in the form of $e_{b_{\alpha\beta}}$ term yields explicit corrections to MAL with influences stemming from correlations. 
The evolution of correlations of the $b_{\alpha\beta}$-type were observed in \cite{uribe2006}.


\section{Conclusion}

\begin{figure}[ht!]
    \begin{center}
        \includegraphics[scale=0.35]{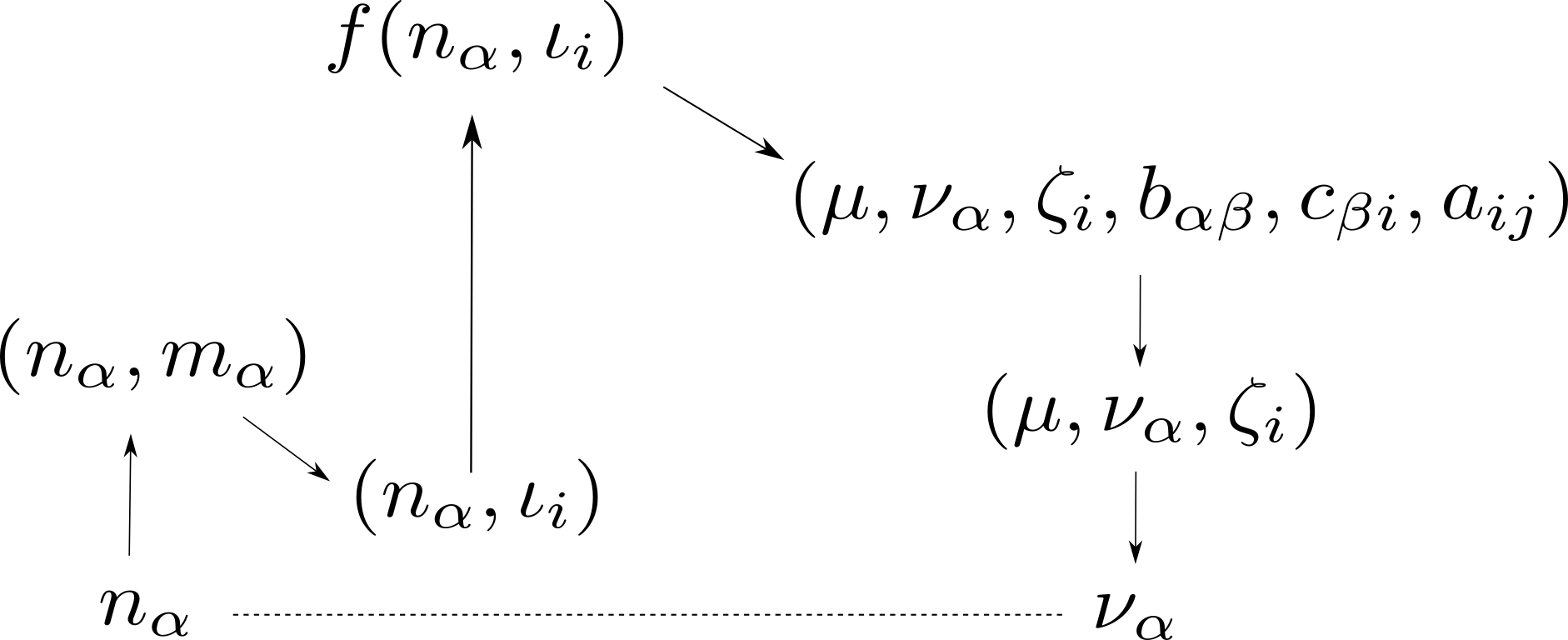}
        \caption{\label{fig.workflow}Levels of description used in this manuscript. Reading from the left bottom corner, we start with the classical MAL dynamics of the compositions $n_\alpha$ and extend it to the $(n_\alpha, \iota_i)$-level, where the rates play the role of state variables as well. This description is then Liouville-lifted to the kinetic theory for $f(\nn,\mathcal{I})$, which is subsequently reduced to the first and second moments. The first moments can be separated and further reduce back to the average compositions $\nu_\alpha$, if necessary. On the way, however, we can observe many phenomena related to the inertial effects and to the effects of fluctuations and their correlations.}
    \end{center}
\end{figure}

Multiscale thermodynamics expresses mathematically the experience collected in  investigations  of  relations among microscopic, mesoscopic, and macroscopic theories of macroscopic systems. In this paper, we present  chemical kinetics as its particular realization. Both multiscale thermodynamics and chemical kinetics become  enriched. The former by providing  its abstract concepts with new physical interpretations and the latter  by acquiring the tools and methods collected in the former. We illustrate  some of the tools  in  investigations of reductions,  fluctuations,  and  coupling  to  mechanics of chemically reacting systems. In bringing thermodynamics and statistical mechanics to the assistance of chemical kinetics, we follow  the pioneering work of Alexander Gorban and Ilya Karlin \cite{GK}.
The various levels of description can be seen in Fig. \ref{fig.workflow}.

Passages from an upper level (a level involving more details) to a lower level (a level involving less details) involve  ignorance of
unimportant details and emergence of important overall features. The essential and the most difficult part of establishing a relation between an upper and a lower level is, of course,  the recognition of what is important and what can be ignored. Inspired by Boltzmann \cite{Boltzmann},  we make the distinction between the fast time evolution that generates unimportant details and the slow time evolution that generates emerging features  passing  to the lower level more pronounced  by breaking the time symmetry and introducing a novel potential (entropy) of nonmechanical origin. The fast time evolution (driven by collisions of particles in the Boltzmann investigation) becomes manifestly distinguished from the slow time evolution (driven by the free motion of particles in the Boltzmann investigation) by different symmetry with respect to the time reversal. The fast time evolution is time irreversible and the slow time evolution is time reversible.

In the context of chemical kinetics, we make the slow time evolution more visible  by making it time reversible and making it driven by energy, while the fast time evolution (and in MAL the complete time evolution) is time irreversible and is driven by entropy. The difference between slow and fast can be expressed in the extended MAL in two sets of parameters: in the rate coefficients (as in classical MAL) and in "reaction mass" (an analog of the particle mass in mechanics).
The extended MAL with reaction rates included in the set of  state variables and the time reversible term in its vector field acquires the full GENERIC structure that has been extracted from mechanical theories. We can then apply  all the tools collected in the multiscale thermodynamics.

In particular, we explore the Liouville lift of the time evolution in a finite-dimensional state space to the time evolution in the infinite-dimensional space of functions on the finite dimensional state  space. The kinetic theory forms of the mass action law that arise in this way bring to chemical kinetics the possibility to take into account the effects of fluctuations. The kinetic theory is then equipped with dissipation, which becomes a Fokker-Planck equation. After the reduction to moments of the distribution function, forming a Grad-like hierarchy, we obtain in Sec. \ref{2nd} evolution equations for the average compositions, reaction rates, tensors of composition-composition, rate-rate, and composition-rate correlations, and entropy.

These correspondences have two implications: i) it is a confirmation of the appropriateness of the considered Liouville lift \eqref{Liu1}; ii) the system \eqref{ncba} indeed represents the kinetic extension of the (at least thermodynamically linearized) classical MAL theory, but containing the influence of fluctuations.

 Still another door opened by the extended MAL is to investigations of the  coupling between chemical kinetics and mechanics.
Specific applications of the initial version of the extended MAL \cite{miroslav-driven}, \cite{grchemkin} notably the problem  of coupling mechanics with chemical reactions (arising, for example, in healing broken bones),  have  been investigating in \cite{klika2013-pre}, \cite{vasek-doblare}, \cite{vk2014-bz}.

\section*{Acknowledgment}
MP and VK were supported by Czech Science Foundation, project no. 20-22092S, and MP also by Charles University Research program No. UNCE/SCI/023.


\begin{thebibliography}{10}

\bibitem{guldberg1879}
C.M. Guldberg and P.~Waage.
\newblock Concerning chemical affinity.
\newblock {\em Erdmann's Journal für Practische Chemie}, 127(69-114), 1879.

\bibitem{MG-CR}
Miroslav Grmela.
\newblock Externally driven macroscopic systems: Dynamics versus
  thermodynamics.
\newblock {\em Journal of Statistical Physics}, 166(2):282--316, 2017.

\bibitem{miroslav-MT}
M.~Grmela.
\newblock Multiscale thermodynamics.
\newblock {\em Entropy}, 23(165), 2021.

\bibitem{renger-fluxes}
D.~R.~Michiel Renger.
\newblock Gradient and generic systems in the space of fluxes, applied to
  reacting particle systems.
\newblock {\em Entropy}, 20(8), 2018.

\bibitem{adv}
M.~Grmela.
\newblock Multiscale equilibrium and nonequilibrium thermodynamics in chemical
  engineering.
\newblock {\em Advances in Chemical Engineering}, 39:76--128, 2010.

\bibitem{miroslav-guide}
Miroslav Grmela.
\newblock {GENERIC} guide to the multiscale dynamics and thermodynamics.
\newblock {\em J. Phys. Commun.}, 2(032001), 2018.

\bibitem{pkg}
Michal Pavelka, V{\' a}clav Klika, and Miroslav Grmela.
\newblock {\em Multiscale Thermo-Dynamics}.
\newblock de Gruyter (Berlin), 2018.

\bibitem{miroslav-ptrsa}
Miroslav Grmela, V{\' a}clav Klika, and Michal Pavelka.
\newblock Gradient and {GENERIC} evolution towards reduced dynamics.
\newblock {\em Phil. Trans.R.Soc.A}, 378(20190472), 2020.

\bibitem{miroslav-driven}
M.~Grmela.
\newblock Thermodynamics of driven systems.
\newblock {\em Phys. Rev. E}, 48:919--930, 1993.

\bibitem{grchemkin}
M.~Grmela.
\newblock Fluctuations in extended mass-action-law dynamics.
\newblock {\em Physica D Nonlinear Phenomena}, 241:976--986, May 2012.

\bibitem{klikajpcb2009coupling}
V{\' a}clav Klika and Franti{\v s}ek Mar{\v s}{\' i}k.
\newblock Coupling effect between mechanical loading and chemical reactions.
\newblock {\em The Journal of Physical Chemistry B}, 113(44):14689--14697,
  2009.

\bibitem{klika2013-pre}
V\'aclav Klika and Miroslav Grmela.
\newblock Coupling between chemical kinetics and mechanics that is both
  nonlinear and compatible with thermodynamics.
\newblock {\em Phys. Rev. E}, 87:012141, Jan 2013.

\bibitem{vasek-remodelling}
V.~Klika and F.~Mar{\v s}{\'i}k.
\newblock A thermodynamic model of bone remodelling: the influence of dynamic
  loading together with biochemical control.
\newblock {\em J Musculoskelet Neuronal Interact}, 10(3):220--230, 2010.

\bibitem{vk2014-bz}
Vaclav Klika and Miroslav Grmela.
\newblock {Mechano-chemical coupling in Belousov-Zhabotinskii reactions}.
\newblock {\em {Journal of chemical physics}}, {140}({12}), {MAR 28} {2014}.

\bibitem{kramers}
H.~A. Kramers.
\newblock Brownian motion in a field of force and the diffusion model of
  chemical reactions.
\newblock {\em Physica}, VII(4), 1940.

\bibitem{Cafel}
Daniel~F. Calef and Peter~G. Wolynes.
\newblock Classical solvent dynamics and electron transfer. 1. {C}ontinuum
  theory.
\newblock {\em J. Phys. Chem.}, 87:3387--3400, 1983.

\bibitem{Boltzmann}
L.B. Gesamtausgabe.
\newblock {\em Ludwig Boltzmann Gesamtausgabe - Collected Works}.
\newblock 1983.

\bibitem{arnoldbook}
V.~I. Arnold.
\newblock {\em Mathematical methods of classical mechanics}.
\newblock Springer, New York, 1989.

\bibitem{GorbanTsym}
Gorban~A. N.
\newblock Detailed balance in micro- and macrokinetics and
  microdistinguishability of macro-processes.
\newblock {\em Results Phys}, 4:142--147, 2014.

\bibitem{pre15}
M.~Pavelka, V.~Klika, and M.~Grmela.
\newblock Time reversal in nonequilibrium thermodynamics.
\newblock {\em Phys. Rev. E}, 90(062131), 2014.

\bibitem{go}
Miroslav Grmela and Hans~Christian \"{O}ttinger.
\newblock Dynamics and thermodynamics of complex fluids. {I}. {D}evelopment of
  a general formalism.
\newblock {\em Phys. Rev. E}, 56:6620--6632, Dec 1997.

\bibitem{og}
Hans~Christian \"Ottinger and Miroslav Grmela.
\newblock Dynamics and thermodynamics of complex fluids. {II}. {I}llustrations
  of a general formalism.
\newblock {\em Phys. Rev. E}, 56:6633--6655, Dec 1997.

\bibitem{hco}
H.C. {\"O}ttinger.
\newblock {\em Beyond Equilibrium Thermodynamics}.
\newblock Wiley, 2005.

\bibitem{dgm}
S.~R. de~Groot and P.~Mazur.
\newblock {\em Non-equilibrium Thermodynamics}.
\newblock Dover Publications, New York, 1984.

\bibitem{Liouville}
J.~Liouville.
\newblock Note sur la theorie de la variation des constants arbitraires.
\newblock {\em J. Math. Appl.}, 3(342), 1838.

\bibitem{koopman}
B.O. Koopman.
\newblock Hamiltonian systems and transformations in {H}ilbert space.
\newblock {\em Proc. Natl. Acad. Sci. USA}, 17(315), 1931.

\bibitem{carleman}
T.~Carleman.
\newblock Application de la th\'{e}orie des \'{e}quations int\'{e}grales
  lin\'{e}aires aux syst`emes d’\'{e}quations diff\'{e}rentielles non
  lin\'{e}aires.
\newblock {\em Acta Math.}, 59(63), 1932.

\bibitem{momentum-Euler}
O{\u g}ul Esen, Miroslav Grmela, Hasan G{\u u}mral, and Michal Pavelka.
\newblock Lifts of symmetric tensors: {F}luids, plasma, and grad hierarchy.
\newblock {\em Entropy}, 21(9):907, 2019.

\bibitem{arnold}
V.I. Arnold.
\newblock Sur la g\'{e}ometrie diff\'{e}rentielle des groupes de lie de
  dimension infini et ses applications dans l'hydrodynamique des fluides
  parfaits.
\newblock {\em Annales de l'institut Fourier}, 16(1):319--361, 1966.

\bibitem{Grad}
H.~Grad.
\newblock {\em Encyclopedia of Physics}, volume~12, chapter Principles of
  Kinetic Theory of Gases.
\newblock Springer-Verlag, 1958.

\bibitem{chapman}
S.~Chapman, T.G. Cowling, D.~Burnett, and C.~Cercignani.
\newblock {\em The Mathematical Theory of Non-uniform Gases: An Account of the
  Kinetic Theory of Viscosity, Thermal Conduction and Diffusion in Gases}.
\newblock Cambridge Mathematical Library. Cambridge University Press, 1990.

\bibitem{struch}
H.~Struchtrup.
\newblock {\em Macroscopic transport equations for rarefied gas flows}.
\newblock Springer, Berlin, Germany, 2005.

\bibitem{Klimontovich1990}
Yu.L. Klimontovich.
\newblock Ito, stratonovich and kinetic forms of stochastic equations.
\newblock {\em Physica A: Statistical Mechanics and its Applications},
  163(2):515--532, 1990.

\bibitem{DynMaxEnt}
V{\' a}clav Klika, Michal Pavelka, Petr V{\' a}gner, and Miroslav Grmela.
\newblock Dynamic maximum entropy reduction.
\newblock {\em Entropy}, 21(715), 2019.

\bibitem{uribe2006}
Carlos~A. Gómez-Uribe and George~C. Verghese.
\newblock Mass fluctuation kinetics: Capturing stochastic effects in systems of
  chemical reactions through coupled mean-variance computations.
\newblock {\em THE JOURNAL OF CHEMICAL PHYSICS}, 126(024109), 2007.

\bibitem{GK}
A.N. Gorban and I.V. Karlin.
\newblock {\em Invariant Manifolds for Physical and Chemical Kinetics}.
\newblock Lecture Notes in Physics. Springer, 2005.

\bibitem{vasek-doblare}
V.~Klika, M.~A. P{\'e}rez, J.~M. Garc{\'i}a-Aznar, F.~Marsik, and
  M.~Doblar{\'e}.
\newblock A coupled mechano-biochemical model for bone adaptation.
\newblock {\em Journal of mathematical biology}, 69(6):1383--1429, 2014.

\bibitem{deDonder}
T.~de~Donder and P.~van Rysselberghe.
\newblock {\em Thermodynamic Theory of Affinity. A Book of Principles}.
\newblock Stanford Univ. Press, 1936.

\bibitem{sieniutycz}
S.~Sieniutycz.
\newblock From a least action principle to mass action law and extended
  affinity.
\newblock {\em Chem. Eng. Sci.}, 42:2697--2711, 1987.

\bibitem{prigogine-tip}
I.~Prigogine.
\newblock {\em Thermodynamics of Irreversible Processes}.
\newblock Thomas, 1955.

\bibitem{om}
L.~Onsager and S.~Machlup.
\newblock Fluctuations and irreversible processes.
\newblock {\em Physical Review}, 91(6):1505--1512, 1953.

\bibitem{gyarmati}
I.~Gyarmati.
\newblock {\em Non-equilibrium thermodynamics: Field theory and variational
  principles}.
\newblock Engineering science library. Springer, 1970.

\bibitem{evans-sde}
L.C. Evans.
\newblock {\em An Introduction to Stochastic Differential Equations}.
\newblock American Mathematical Society, 2012.

\bibitem{hco-jnet2020-I}
Hans~Christian {\" O}ttinger, Mark~A. Peletier, and Alberto Montefusco.
\newblock A framework of nonequilibrium statistical mechanics. i. role and
  types of fluctuations.
\newblock {\em Journal of Non-Equilibrium Thermodynamics}, 46(1):1--13, 2021.

\end{thebibliography}

\appendix

\section{Properties of Solutions to Eq.(\ref{dnwdt})}\label{sred1}

Now we begin  to investigate the properties of solutions to (\ref{dnwdt}). We anticipate that with an appropriate choice of $\phi$ and $\vartheta$ solutions to (\ref{dnwdt}) reveal patterns representing solutions to equations governing the time evolution in
less detailed descriptions  of chemically reacting systems. In particular, we anticipate to recover in this way the equilibrium thermodynamic characterization of the chemical equilibrium and the classical mass action law.

In general, the problem of solving qualitatively the time evolution equations like Eq.(\ref{dnwdt}) is the problem of recognizing a pattern in the phase portrait corresponding to (\ref{dnwdt}). We recall that the phase portrait is a collection of solutions to (\ref{dnwdt}) passing through all $(\nn,\mathcal{I})$ for an ensemble of systems corresponding to a  set of material parameters. In Eq.(\ref{dnwdt}) the material parameters are the two potentials $\phi$ and $\vartheta$.

Due to the antisymmetry of the Poisson bracket,  the generating potential $\phi$ is conserved in the time evolution governed by the Hamiltonian version of (\ref{dnwdt}) (i.e. Eq.(\ref{dnwdt}) without the last term on its right hand side).  However, also other potentials can be conserved during this Hamiltonian time evolution because the Poisson bracket (\ref{pb2}) is, in general,  degenerate in the sense that there exist potentials $c$ for which $\{a,c\}=0, \forall a$. Such potentials are called Casimirs. Indeed, the conservation follows directly from  (\ref{pb1}), $\dot{c}=\{c,\phi\}=0$.

Regarding solutions to the complete equation  (\ref{dnwdt})  with the last term involving the dissipation potential $\vartheta$ satisfying (\ref{disspot}),  we note that
\begin{equation}\label{pb5}
\dot{\phi}=-\langle\mathcal{I}^*,\vartheta_{\mathcal{I}^*}\rangle < 0,
\end{equation}
because  $\dot{\phi}=0$ in the Hamiltonian time evolution and the inequality on the right hand side of (\ref{pb5}) is a direct consequence of (\ref{disspot}).
The relation (\ref{pb5}) together with the convexity of $\phi$ implies  that the thermodynamic  potential $\phi$ plays in the time evolution governed by (\ref{dnwdt}) the role of the Lyapunov function  associated with the approach, as $t\rightarrow\infty$, to states at which $\mathcal{I}^*=0$.

\subsection{Extended mass action law $\longrightarrow$ Classical mass action law}\label{clasmas}

To show that the disappearance of the conjugate fluxes $\mathcal{I}^*$ implies the  chemical equilibrium known from the equilibrium thermodynamics, we have to find more details about solutions to (\ref{dnwdt}).
We also identify the potentials $\phi$ and $\vartheta$ for which the mass action law reduced from the extended mass action law (\ref{dnwdt}) becomes the classical Guldberg-Waage mass action law. To be able to do it, we need to make  approximations. We assume that their validity will be guaranteed by an appropriate choice (made on the basis of physical considerations) of the potential $\phi$ and $\vartheta$ that remain so far in Eq.(\ref{dnwdt}) undetermined.

First,  we note  that in the case of small inertia and strong dissipation, the right hand side of the second equation in (\ref{dnwdt}) is larger than the right hand side of the first equation. Indeed, the small inertia means that $\mathcal{I}^*\sim m \mathcal{I}$, where the parameter $m$, called a reaction mass,  (an analogue of mass in the context of mechanics) is small. This means that the first  term on the right hand side of the second equation in (\ref{dnwdt}) is proportional to the large parameter $\frac{1}{m}$ and the second term $\vartheta_{\mathcal{I}^*}\sim\lambda \mathcal{I}^*$, where $\lambda$  is large (of the same order  as $\frac{1}{m}$ ) if the dissipation (friction in mechanics) is large.

Consequently, the second equation in (\ref{dnwdt}) can be approximated by
$-\gamma^T\nn^*-\vartheta_{\mathcal{I}^*}=0$
which can also be written as
\begin{equation}\label{red1}
\left[\psi_{\mathcal{I}^*}(\mathcal{I}^*,(\mathcal{I}^*)^{\dag}; \nn,\mathcal{I})\right]_{(\mathcal{I}^*)^{\dag}=-\gamma^T\nn^*}=0,
\end{equation}
with
\begin{equation}\label{Psi}
\psi(\mathcal{I}^*,(\mathcal{I}^*)^{\dag}; \nn,\mathcal{I})=-\vartheta(\mathcal{I}^*; \nn,\mathcal{I})+\langle\mathcal{I}^*,(\mathcal{I}^*)^{\dag}\rangle,
\end{equation}
where the upper index $\dag$ denotes conjugation with respect to the dissipation potential $\vartheta$ (i.e. $(\mathcal{I}^*)^{\dag}=\vartheta_{\mathcal{I}^*}$). We note that
\begin{equation}\label{Idag}
\mathcal{I}^*=\vartheta^{\dag}_{(\mathcal{I}^*)^{\dag}}((\mathcal{I}^*)^{\dag};\nn,\mathcal{I}),
\end{equation}
where
\begin{equation}\label{IIdag}
\vartheta^{\dag}((\mathcal{I}^*)^{\dag};\nn,\mathcal{I})=\psi(\widehat{\mathcal{I}^*}((\mathcal{I}^*)^{\dag});(\mathcal{I}^*)^{\dag},\nn,\mathcal{I})
\end{equation}
is the conjugate dissipation potential and  $\widehat{\mathcal{I}^*}((\mathcal{I}^*)^{\dag},\nn,\mathcal{I})$ is a solution to (\ref{red1}).

If we now insert (\ref{Idag}) into the first equation in (\ref{dnwdt}) we obtain
\begin{equation}\label{class}
\frac{d\nn}{dt}=\gamma \left[\vartheta^{\dag}_{(\mathcal{I}^*)^{\dag}}\right]_{(\mathcal{I}^*)^{\dag}=-\gamma^T\nn^*}.
\end{equation}
Using the   simplified notation
\begin{equation}\label{Xi}
\xi(\nn^*,\nn)=[\vartheta^{\dag}((\mathcal{I}^*)^{\dag};\nn,\mathcal{I})]_{(\mathcal{I}^*)^{\dag}=-\gamma^T\nn^*},
\end{equation}
Eq.(\ref{class}) becomes
\begin{equation}\label{class1}
\frac{d\nn}{dt}=-\xi_{\nn^*}(\nn^*,\nn,\mathcal{I}),
\end{equation}
which is  the   classical mass action law.

Now we are in position to show  that:

(i)
 States for which $\mathcal{I}^*=0$ (i.e. the states reached as $t\rightarrow \infty$ in the time evolution governed by (\ref{dnwdt})) are the chemical equilibrium states defined in the classical equilibrium thermodynamics (i.e. the states for which the chemical affinities equal zero).

 (ii) The potentials $\phi$ and $\vartheta$  appearing in (\ref{dnwdt}) for which the reduced mass action law (\ref{class}) becomes the classical Guldberg-Waage mass action law.

 To prove the first statement, we note that due to the properties (\ref{disspot}) of $\vartheta$, $\mathcal{I}^*=0$ if and only if $(\mathcal{I}^*)^{\dag}=0$. Since $(\mathcal{I}^*)^{\dag}= -\gamma^T\nn^*$, the equilibrium is reached when $\gamma^T\nn^*=0$. However, $\gamma^T\nn^* $ are the chemical affinities introduced in the classical equilibrium thermodynamics. We have thus proven that the state approached in the time evolution governed by the extended mass action law (\ref{dnwdt}) are the chemical equilibrium states introduced  in the classical equilibrium thermodynamics. Moreover, we have shown that the long-time behavior of solutions to the extended mass action law (\ref{dnwdt}) is well approximated by solutions to the classical mass action law (\ref{class}).

 It remains  to identify the potentials $\phi(\nn,\mathcal{I})$ and $\vartheta$ for which (\ref{class}) becomes the classical Guldberg-Waage mass action law. 
 One can directly verify that the standard choice of energy \eqref{Phicond} together with the following dissipation potential
 \begin{equation*}
   \vartheta=L \sum_j \mathcal{I}^*_j \mathrm{arcsinh}\left(\lambda_j(\nn)\mathcal{I}^*_j\right)+\frac{1}{\lambda_j(\nn)} \left(1-\sqrt{1+(\lambda_j(\nn) \mathcal{I}^*_j)^2}\right)\approx L \frac{\lambda_j}{2} (\mathcal{I}^*_j)^2,
 \end{equation*}
yield exactly the reduction to the classical mass action law. To see this explicitly, we rewrite MAL \eqref{eq:classMAL} into a different form
\begin{equation*} 
  \frac{d\nu_{\alpha}}{dt} = \sum_j \gamma_{\alpha j} 2 \overrightarrow{k}_j e^{\frac{1}{2}\sum_\beta \gamma_{\beta j} (Q_\beta+1)} \left(\prod_\beta \nu_\beta^{\nu_{\beta j}+\mu_{\beta i}}\right)^{-1/2}  \sinh (\sum_\beta \gamma_{j\beta} \nu_\beta^*),
\end{equation*}
where $\nu_\beta^*=s_{\nu_\beta}=\frac{1}{T} e_{\nu_\beta}$. One can see that with the appropriate choice of $L, \lambda_j$ one can obtain exactly the classical MAL after the reduction.

Summing up, we have demonstrated that the three levels of investigation of chemically reacting systems, namely the extended mass action law (\ref{dnwdt}), the classical mass action law (\ref{class}) and the level of the classical equilibrium thermodynamics,  are mutually compatible. The possibility and advantage of formulating the Guldberg-Waage mass action law in the form (\ref{class}), (\ref{Phicond}), (\ref{Thetacond}) has appeared gradually in \cite{deDonder} - \cite{sieniutycz}, \cite{miroslav-driven}.

The level represented by (\ref{dnwdt}) involves the most details, the classical equilibrium thermodynamics the least details. In the next section, we show that  the compatibility between (\ref{dnwdt}) and (\ref{class}) provides a basis for introducing rate thermodynamics of chemically reacting systems.

\subsection{Rate thermodynamics }\label{rateth}

Our attention in the investigation of solutions to (\ref{dnwdt}) was put in the previous section to the large time behavior under the assumption  of small inertia and large dissipation. The asymptotic time evolution has been found to be well approximated by the first equation in which the flux $\mathcal{I}^*$ is expressed in terms of $\nn^*$. In this section, we put into focus the short time behavior of systems that can be externally driven and do not, in general,  approach chemical equilibrium. We shall see that with this focus, Eq.(\ref{dnwdt}) can be well approximated by the second equation describing the time evolution of fluxes and forces.

It is useful to introduce first a simplifying notation that has also the advantage of bringing the analysis of solutions to (\ref{dnwdt}) closer to some investigations in the classical nonequilibrium thermodynamics. In this section we shall use $(\JJ,\xx)$ (fluxes and forces) instead of the $(\nn,\mathcal{I})$ state variables:
\begin{eqnarray}\label{not}
\JJ&=&\mathcal{I}^*,\nonumber \\
\xx&=&-\gamma^T \nn^*.
\end{eqnarray}

With this notation, let us estimate the reducing evolution, the fast evolution on a small time scale. Let us assume that this fast transition is fast when compared to the evolution of $\frac{\partial \nn}{\partial t}$ and thus $\nn$ and $\nn^*$ is approximately constant during this evolution entailing $\xx$ begin a constant as well. For clarity, we shall denote it $\overline{\xx}$.
Then, the second equation in (\ref{dnwdt}) becomes
\begin{equation}\label{rt1}
  \frac{\partial J_i}{\partial t}=\frac{\partial \mathcal{I}^*}{\partial t}=\mathbb{G}_{ij}(\overline{x}_j-\vartheta_{\mathcal{I}_j^*}),
\end{equation}
where
\begin{equation}\label{Gx}
\mathbb{G}_{ij}=\phi_{\iota_i \iota_j},
\end{equation}
where $\overline{\xx}$ is undetermined at this point but slaved to $\JJ$. If we denote $\psi$ the Legendre transform of $\vartheta$:
\begin{equation}\label{Psirt}
\psi(\JJ,\xx; \nn,\mathcal{I})=-\vartheta(\JJ; \nn,\mathcal{I})+\langle\JJ,\xx\rangle,
\end{equation}
we may rewrite the fast evolution equation \eqref{rt1} as
\begin{equation}\label{rt2}
  \frac{\partial J_i}{\partial t}=\mathbb{G}_{ij} \psi_{J_j}(\JJ,\overline{\xx};\overline{\nn},\mathcal{I}).
\end{equation}

We now give  Eq.(\ref{rt2}) the role of  a  fundamental equation in  a theory in which the only state variable is the flux $\JJ$ and $\psi$ is the potential driving its evolution. 
We call such theory   \textit{rate thermodynamics} and the potential
$\vartheta(\JJ;\overline{\nn},\mathcal{I})$   \textit{rate free energy}. The thermodynamic state variables  $(\overline{\nn}, \mathcal{I})$ play  in the rate thermodynamics the role of  parameters. To simplify the notation, we shall omit to write them in the rest of this section. The convexity of the free energy $\phi$ implies that $\mathbb{G}$ is a positive definite operator. The vector $\overline{\xx}$ appearing in (\ref{rt1}), (\ref{rt2}) is in the rate thermodynamics an unspecified parameter expressing mathematically the thermodynamic forces, in particular the chemical affinity $\gamma^T\nn^*$.

Now we begin to investigate the properties of solutions to (\ref{rt2}). First, we note that
\begin{equation}\label{rt3}
\frac{\partial \psi}{\partial t}=\langle\psi_{\JJ}, \mathbb{G}\psi_{\JJ}\rangle > 0
\end{equation}
which, together with the requirements (\ref{disspot}), implies that $-\psi$ plays the role of the Lyapunov function for the $t\rightarrow\infty$ approach of solutions to (\ref{rt2}) to $\widehat{\JJ}(\overline{\xx})$ that is a solution to
\begin{equation}\label{rt4}
\psi_{\JJ}(\JJ,\overline{\xx})=0
\end{equation}
From the physical point of view, we can interpret (\ref{rt4}) as a minimization of the rate free energy $\vartheta$ with the constraint $\JJ$ and $\overline{\xx}$ playing the role of the Lagrange multiplier. Alternatively, we can also see (\ref{rt4}) as the first step in making a Legendre transformation from $\vartheta(\JJ)$ to $\vartheta^{\dag}(\overline{\xx})=\psi(\widehat{\JJ}(\overline{\xx}),\overline{\xx})$. The rate free energy  $\vartheta$ plays in the rate thermodynamics the role that the free energy $\phi$ plays in thermodynamics. Moreover, we have seen that the rate free energy plays in the extended mass action law the role of the dissipation potential and its conjugate with (\ref{Gx}) is the dissipation potential $\xi$ appearing in the classical mass action law (\ref{class}).

We have seen that with the choice (\ref{Gx}) the governing equation (\ref{rt2}) of the rate thermodynamics describes an approach from the extended to the classical mass action law. Giving the rate thermodynamics the role of a bond binding the extended and the classical mass action law is a novel contribution to the thermodynamics of fluxes and forces that emerged in the classical nonequilibrium thermodynamics in Refs.\cite{prigogine-tip}, \cite{om}, \cite{gyarmati}.

Before leaving this section, we emphasize that the rate thermodynamics is applicable to externally driven systems to which the classical thermodynamics does not apply. The driving forces enter in the vector $\overline{\xx}$. The potential $\vartheta^{\dag}(\overline{\xx})$ or its Legendre transform $\vartheta(\JJ)$ play in the rate thermodynamics the role that the free energy $\phi(\nn)$ plays in thermodynamics. If the force $\overline{\xx}$ is chosen to be the chemical affinity then the rate thermodynamic potential becomes the dissipation potential $\xi(\nn^*,\nn)$ entering the mass action law (i.e.  the free energy production $\frac{\partial \phi}{\partial t}=\langle\nn^*,\xi_{\nn^*}\rangle$).

\section{BGK dynamics}\label{sec.BGK}
The BGK irreversible dynamics, see e.g. \cite{GK}, is a popular alternative to the Boltzmann collision integral in the evolution of the distribution function $f$, 
\begin{equation}\label{eq.BGK.irr}
    (\partial_t f)_{irr} = -\frac{1}{\tau}(f - f_{eq}).
\end{equation}
On the other hand, irreversible evolution within GENERIC is prescribed as gradient dynamics, i.e. as the derivative of a dissipation potential w.r.t $f^*$. Is BGK compatible with GENERIC? It indeed is, but we shall first carry out a few calculations. 

\subsection{BGK within GENERIC}
Consider the Boltzmann entropy $S(f) = -k_B \int d\rr\int d\pp f (ln(h^3f)-1)$. The equilibrium distribution is obtained by maximization of the entropy while keeping the equilibrium state variables as constraints, 
\begin{equation}
    \frac{\delta}{\delta f}\left( -S + E^*\int d\rr\int d\pp e(\rr,\pp) f + N^* \int d\rr\int d\pp f\right) = 0,
\end{equation}
which leads to the MaxEnt estimate of the distribution function,
\begin{equation}
    f_{eq}(E^*,N^*) = \frac{1}{h^3}e^{-E^*e/k_B}e^{-N^*/k_B}.
\end{equation}
The conjugate distribution function $f^* = S_f$ evaluated at the MaxEnt estimate then becomes
\begin{equation}
    f^*_{eq} = E^* e + N^*.
\end{equation}

The thermodynamic force driving the evolution towards equilibrium in the BGK approximation, 
\begin{equation}
    X = \frac{f^*-f^*_{eq}}{2k_B},
\end{equation}
is constructed as the difference between the actual conjugate distribution function and the equilibrium value. 
The dissipation potential generating the BGK irreversible evolution is then analogical to the dissipation potential generating chemical kinetics \cite{grchemkin}, 
\begin{equation}
    \Xi(f,f^*,N^*,E^*) = \frac{4k_B}{\tau}\sqrt{f \cdot f_{eq}} \cosh(X(f^*,E^*,N^*)).
\end{equation}
Indeed, the gradient dynamics $(\partial_t f)_{irr} = \Xi_{f^*}|_{f^* = S_f}$ then becomes equivalent with the BGK evolution \eqref{eq.BGK.irr}.

\subsection{BGK-like dissipation in chemical kinetics} \label{BGKdis}
Another option how to ensure the approach of the distribution function to the equilibrium value is the Hamiltonian and BGK dynamics, 
\begin{eqnarray}\label{eq.BGK}
\frac{\partial f}{\partial t} &=&-\frac{\partial}{\partial n_{\alpha}}\left(f\gamma_{\alpha j}\frac{\partial f^*}{\partial \iota_j}\right)+\frac{\partial}{\partial \iota_j}\left(f\gamma_{j\alpha}^T\frac{\partial f^*}{\partial n_{\alpha}}\right)
    + \frac{\delta \Xi^{(BGK)}}{\delta f^*}\Big|_{f^*=S_f}\nonumber\\ 
    &=& -\frac{\partial}{\partial n_{\alpha}}\left(f\gamma_{\alpha j}\frac{\partial f^*}{\partial \iota_j}\right)+\frac{\partial}{\partial \iota_j}\left(f\gamma_{j\alpha}^T\frac{\partial f^*}{\partial n_{\alpha}}\right)
    -\frac{1}{\tau}(f-f_{eq}),
\end{eqnarray}
where the BGK dissipation potential reads
\begin{equation}
    \Xi^{(BGK)}  = \frac{4k_B}{\tau}\sqrt{f \cdot f_{eq}} \cosh\left(\frac{f^*-f^*_{eq}}{2k_B}\right)
\end{equation}
and where $f^*_{eq}$ is the equilibrium conjugate distribution function, see Appendix \ref{sec.BGK}.

\begin{subequations}\label{eq.BGK.cons}
Conservation of energy $E= \int d\rr\int d\pp ef$ in the Hamiltonian part is automatic due to the antisymmetry of the bracket. The irreversible part conserves energy as well, since
\begin{equation}
    \dot{E} = \int d\rr\int d\pp E_f \partial_t f = \int d\rr\int d\pp e  \frac{1}{\tau} (f-f_{eq}) = E - E = 0.
\end{equation}
Similarly, the normalization $N = \int d\rr\int d\pp f$ is conserved as well due to 
\begin{equation}
    \dot{N} = \int d\rr\int d\pp N_f \partial_t f = \int d\rr\int d\pp  \frac{1}{\tau} (f-f_{eq}) = N - N = 0.
\end{equation}
\end{subequations}

The dissipation potential of course ensures the non-negativity of the entropy production, 
\begin{align}
    \dot{S} &= \int d\rr\int d\pp S_f \frac{2}{\tau}\sqrt{f \cdot f_{eq}}\sinh\left(\frac{f^*-f^*_{eq}}{2k_B}\right)\nonumber\\
     &= \int d\rr\int d\pp (f^*-f^*_{eq})\frac{2}{\tau}\sqrt{f \cdot f_{eq}}\sinh\left(\frac{f^*-f^*_{eq}}{2k_B}\right)\geq 0, 
\end{align}
where the last equality follows from the conservation properties \eqref{eq.BGK.cons}.

\subsection{BGK-like dissipation}
Let us, for simplicity start with the zero-th and first moments only, $\xx = (\mu, \nu_\alpha, \zeta_i)$, letting the energy depend only on these moments. 

The reversible part of the evolution equations is then obtained as a part of Eqs. \eqref{ncba},
\begin{subequations}\label{eqs.first.rev}
\begin{eqnarray}
 \left(\frac{d\mu}{dt}\right)_{rev}&=&0, \\
 \left(\frac{d\nu_{\alpha}}{dt}\right)_{rev}&=&\gamma_{\alpha j}\mu e_{\zeta_j}\\
 \left(\frac{d\zeta_j}{dt}\right)_{rev}&=&-\gamma_{j\alpha}^T\mu e_{\nu_{\alpha}}.
\end{eqnarray}
\end{subequations}
The irreversible evolution is obtained by projection of the BGK irreversible part \eqref{eq.BGK}, 
\begin{subequations}
\begin{eqnarray}
 \left(\frac{d\mu}{dt}\right)_{irr}&=&0, \\
    \left(\frac{d\nu_{\alpha}}{dt}\right)_{irr}&=& - \frac{\nu_\alpha-\nu_{\alpha,eq}}{\tau}\\
 \left(\frac{d\zeta_j}{dt}\right)_{irr}&=&- \frac{\zeta_j}{\tau}.
\end{eqnarray}
    where we used that $\mu=\mu_{eq}$ and $\zzeta_{eq} = 0$. The former equality comes from the normalization of $f$ while the latter from the odd parity of $\zzeta$.
\end{subequations}
Altogether, the equations for the first moments are the sum of the reversible and irreversible evolutions,
\begin{subequations}\label{eq.first}
\begin{eqnarray}
 \frac{d\mu}{dt}&=&0, \\
 \frac{d\nu_{\alpha}}{dt}&=&\gamma_{\alpha j}\mu e_{\zeta_j} - \frac{\nu_\alpha-\nu_{\alpha,eq}}{\tau}\\
 \frac{d\zeta_j}{dt}&=&-\gamma_{j\alpha}^T\mu e_{\nu_{\alpha}}- \frac{\zeta_j}{\tau}.
\end{eqnarray}
\end{subequations}
These equations express chemical kinetic with inertial effects. 

In order to write the equations in a closed form, we have to supply to an energy functional. The energy has two parts, one is the usual chemical energy that depends on the composition of the mixture while the other is a part that depends on the actual reaction rates. The latter part expresses the here proposed inertial effects. For instance in \cite{Cafel} it was shown that the solvent can provide some inertia to the chemical kinetics due to the time-dependent evolution of dielectric polarization. However, only the purely irreversible MAL has been derived in \cite{Cafel} from the kinetic theory, so in this paper we go beyond because we keep the inertial contribution. As we do not wish to be restricted to any particular mechanism (e.g. polarization relaxation), we assume a generally convex quadratic contribution to the energy from the reaction rates. The overall energy then reads
\begin{equation}
 e = e(\nnu)_{chem} + \frac{1}{2}K_{ij}\zeta_i \zeta_j,
\end{equation}
where the first part is the classical chemical energy while the latter is the kinetic contribution due to the fluxes with a symmetric positive definite matrix of effective masses $K_{ij}$. 

Let us now discuss the relation of Eqs. \eqref{eq.first} with the usual MAL. Assuming strong dissipation, $\tau \ll 1$, we can use the asymptotic expansion
\begin{equation}
    \nu_\alpha = \nu_{\alpha}^{(0)} + \tau \nu_\alpha^{(1)} + \dots
    \quad\mbox{and}\quad
    \zeta_i = \zeta_{i}^{(0)} + \tau \zeta_i^{(1)} + \dots,
\end{equation}
and Eqs. \eqref{eq.first} then become
\begin{subequations}\label{eqs.first.asymp}
\begin{eqnarray}
    \frac{d\nu^{(0)}_{\alpha}}{dt} + \tau \frac{d\nu^{(1)}_{\alpha}}{dt}+\dots&=&\gamma_{\alpha j}\mu e_{\zeta_j}|_{(\zzeta^{(0)} + \tau\zzeta^{(1)}+\dots)} \nonumber\\
    &&- \frac{\nu^{(0)}_\alpha + \tau \nu^{(1)}_\alpha + \dots -\nu_{\alpha,eq}}{\tau}\\
    \frac{d\zeta^{(0)}_j}{dt} + \tau \frac{d\zeta^{(1)}_j}{dt} + \dots&=&-\gamma_{j\alpha}^T\mu e_{\nu_{\alpha}}|_{(\nnu^{(0)} + \tau \nnu^{(1)}+\dots)}\nonumber\\
    &&- \frac{\zeta^{(0)}_j + \tau \zeta^{(1)}_j + \dots}{\tau}.
\end{eqnarray}
\end{subequations}
From the terms of order $\tau^{-1}$ we can see that 
\begin{equation}
\zzeta^{(0)} = 0 \qquad\mbox{and}\qquad \nnu^{(0)}=\nnu_{eq}. 
\end{equation}
From the terms of order $\tau^0$ we find that 
\begin{equation}
    \dot{\nu}^{(0)}_\alpha = -\nu^{(1)}_\alpha
    \qquad\mbox{and}\qquad
    0 = -\gamma^T_{j\alpha}\mu e_{\nu_\alpha}|_{\nnu_{eq}} - \zeta^{(1)}_j.
\end{equation}
From the terms of order $\tau^1$ we obtain that 
\begin{equation}
    \dot{\nu}^{(1)}_\alpha = \gamma_{\alpha j}\mu K_{jl}\zeta^{(1)}_l
    = -\gamma_{\alpha j}\mu^2 K_{jl} \gamma^T_{l \beta} e_{\nu_\beta}|_{\nnu_{eq}}.
\end{equation}
Finally, the approximate evolution equation for $\nnu$ reads (using the terms of order up to $\tau^1$)
\begin{equation}
    \dot{\nu}_\alpha = -\frac{\nu_\alpha - \nu_{\alpha,eq}}{\tau} 
    -\tau \gamma_{\alpha j}\mu^2 K_{jl} \gamma^T_{l \beta} e_{\nu_\beta}|_{\nnu_{eq}} + \mathcal{O}(\tau^2).
\end{equation}
The first term on the right hand side is the usual MAL while the second term is of higher order in $\tau$ and represents the effect of the inertial terms. If, for instance, $K_{ij} = 0$, or if $\tau$ is negligible, the second term can be neglected, but generally it contributes to the evolution of the chemical composition.

\section{From the noise to the Fokker-Planck equation}\label{sec.Ito}
For the reader's convenience, we include some details on the passage from dynamics \eqref{eq.nI.noise} to the Fokker-Planck equation \eqref{eq.FP}. 
Let us assume some state variables $\xx$ with a stochastic differential equation
\begin{equation}
    dx^i = \mu^i dt + B^{ij} d W_j(t), 
\end{equation}
where $\mu^i$ is a drift term, $B^{ij}$ is the amplitude of the noise, and $W_j$ is an increment of the Wiener process. If we interpret this equation in the Ito sense \cite{evans-sde}, we can employ the Ito lemma, stating that for any smooth function function $a(\xx)$ 
\begin{equation}
    da = \frac{\partial a}{\partial x^i} \mu^i dt + \frac{1}{2}\frac{\partial^2 a}{\partial x^i \partial x^j}B^{ik}dW_k B^{jl}dW_l.
\end{equation}
The square of the Wiener process increment is obviously of the order of $dt$, $\langle W_k W_l\rangle = \delta_{kl}dt$. Taking the average value of the increment, we obtain
\begin{equation}
    \langle da \rangle= \frac{\partial a}{\partial x^i} \mu^i dt + \frac{1}{2}\frac{\partial^2 a}{\partial x^i \partial x^j}B^{ik}\delta_{kl} B^{jl}dt.
\end{equation}

Then we consider a functional $A=\int d\xx a f$, $f$ playing the role of the distribution function. 
Dynamics of $A$ can be constructed in two ways, 
\begin{equation}
    dA = \int d\xx f \langle da \rangle
    \quad\mbox{and}\quad
    dA = \int d\xx a df.
\end{equation}
We have the former and by casting it into the latter form, 
\begin{align}
    \dot{A}=\frac{dA}{dt} &= \int d\xx \left(\frac{\partial a}{\partial x^i} \mu^i + \frac{1}{2}\frac{\partial^2 a}{\partial x^i \partial x^j}B^{ik}\delta_{kl} B^{jl}\right) f(\xx)\nonumber\\
    &= \int d\xx a(\xx) \left(-\frac{\partial \mu^i f}{\partial x^i} + \frac{1}{2}\frac{\partial^2 B^{ik}\delta_{kl} B^{jl} f}{\partial x^i\partial x^j} \right),
\end{align}
we can read the Fokker-Planck equation
\begin{equation}\label{eq.FP.gen}
    \partial_t f = -\frac{\partial \mu^i f}{\partial x^i} + \frac{1}{2}\frac{\partial^2 B^{ik}\delta_{kl} B^{jl} f}{\partial x^i\partial x^j}.
\end{equation}

On the other hand, the Liouville lift within GENERIC \eqref{eq.Genlift} leads to 
\begin{equation}
    \partial_t = -\frac{\partial}{\partial x^i}\left(L^{ij} \frac{\partial E}{\partial x^j}\right) 
    -\frac{\partial}{\partial x^i}\left(M^{ij} \frac{\partial S}{\partial x^j}\right) 
    +\frac{\partial}{\partial x^i}\left(k_B M^{ij} \frac{\partial f}{\partial x^j}\right),
\end{equation}
where $E(\xx)$ and $S(\xx)$ are the energy and entropy in terms of the state variables $\xx$. Note that the first two terms on the right hand side represent the drift term while the third term corresponds to the noise term in the Fokker-Planck equation \eqref{eq.FP.gen}. By comparison of the noise terms, we obtain the fluctuation-dissipation theorem $2k_B \mathbf{M} = \mathbf{B}\mathbf{B}^T$, see also \cite{hco}. Note that the constantness of $M^{ij}$ is necessary for the direct comparison and if it is not constant, the Ito interpretation of the underlying stochastic differential equations itself comes into question \cite{hco-jnet2020-I}.

\end{document}